\documentclass[%
 reprint,
superscriptaddress,
showpacs,
 amsmath,amssymb,
 aps,
floatfix,
]{revtex4-1}

\usepackage{rotating}
\usepackage{stackengine}
\usepackage{graphicx}
\usepackage{color}
\usepackage{verbatim}
\usepackage{subfigure}
\usepackage{tikz}
\usepackage{dsfont}
\usepackage{algpseudocode}
\usepackage{algorithm}

\definecolor{Blue}{rgb}{0,0,1}
\definecolor{Red}{rgb}{1,0,0}
\definecolor{Black}{rgb}{0,0,0}

\usepackage[colorlinks=true,linkcolor=blue,citecolor=blue,urlcolor=blue]{hyperref}

\definecolor{royalblue}{HTML}{4169e1}
\newcommand{\sss}{\scriptscriptstyle} 
\newcommand{\MI}[1]{{\color{Black} #1}}
\newcommand{\MW}[1]{{\color{Black} #1}}
\newcommand{\DB}[1]{{\color{Black} #1}}
\newcommand{\DBR}[1]{{\color{Black} #1}}
\newcommand{\MII}[1]{{\color{Black} #1}}

\begin{document}

\preprint{APS/123-QED}

\title{Metrological detection of entanglement generated by non-Gaussian operations} 

\author{David Barral} 
\email{david.barral@lkb.ens.fr}
\affiliation{Laboratoire Kastler Brossel, Sorbonne Université, CNRS, ENS-PSL Research University, Collège de France, 4 place Jussieu, F-75252 Paris, France} 
\author{Mathieu Isoard} 
\affiliation{Laboratoire Kastler Brossel, Sorbonne Université, CNRS, ENS-PSL Research University, Collège de France, 4 place Jussieu, F-75252 Paris, France}
\author{Giacomo Sorelli} 
\affiliation{Laboratoire Kastler Brossel, Sorbonne Université, CNRS, ENS-PSL Research University, Collège de France, 4 place Jussieu, F-75252 Paris, France}
\affiliation{Fraunhofer IOSB, Ettlingen, Fraunhofer Institute of Optronics, System Technologies and Image Exploitation, Gutleuthausstr. 1, 76275 Ettlingen, Germany}
\author{Manuel Gessner} 
\affiliation{
Departamento de Física Teórica, IFIC, Universidad de Valencia-CSIC, C/ Dr. Moliner 50, Burjassot, Valencia  46100, Spain}
\author{Nicolas Treps} 
\affiliation{Laboratoire Kastler Brossel, Sorbonne Université, CNRS, ENS-PSL Research University, Collège de France, 4 place Jussieu, F-75252 Paris, France}
\author{Mattia Walschaers}
\email{mattia.walschaers@lkb.upmc.fr}
\affiliation{Laboratoire Kastler Brossel, Sorbonne Université, CNRS, ENS-PSL Research University, Collège de France, 4 place Jussieu, F-75252 Paris, France}

\begin{abstract}

Entanglement and non-Gaussianity are physical resources \DBR{that are} essential for a large number of quantum-optics protocols. Non-Gaussian entanglement is indispensable for quantum-computing advantage and outperforms its Gaussian counterparts in a number of quantum-information protocols. The characterization of non-Gaussian entanglement is a critical matter as it is in general highly demanding in terms of resources. We propose a simple protocol based on the Fisher information for witnessing entanglement in an important class of non-Gaussian entangled states: photon-subtracted states. We demonstrate that our protocol is relevant for the detection of non-Gaussian entanglement generated by multiple photon-subtraction and that it is experimentally feasible through homodyne detection.
\end{abstract}

\date{December 5, 2023}
\maketitle 

\section{Introduction}


Entanglement is considered one of the most striking breakthroughs of 20th century science. The gedanken experiment proposed by Einstein, Podolsky and Rosen in 1935 \cite{Einstein1935} pointed out the notion of inseparability of a state composed by two quantum particles spatially distanced with maximally correlated momenta and maximally anti-correlated positions. Nowadays, entanglement stands as a physical resource underpinning most of current development in quantum technologies \cite{Acin2018}. The efficient detection and measurement of entanglement is a very active area of quantum physics \cite{Horodecki2009}, being far from simple especially for continuous variable (CV) systems which involve physical quantities with a continuous spectrum of values \cite{Braunstein2005}.


Multimode squeezed states of light are the cornerstone of CV quantum networks \cite{Larsen2019, Asavanant2019}. They exhibit Gaussian statistics and their entanglement properties are completely specified by their covariance matrix. Criteria and witnesses for this Gaussian entanglement have been proposed and tested for decades \DBR{\cite{Duan2000, Simon2000, Giovanetti2003, vanLoock2003,Abiuso2021}}. However, Gaussian entanglement can always be undone with passive linear optics, a phenomenon generally refereed to as passive separabiliy \cite{Walschaers2017}. It was recently found that one requires states that are not passively separable as a resource for a quantum computational advantage \cite{Chabaud2022}. Because all Gaussian states are passively separable, we can always find mode bases in which the covariance matrix of the state will not show any direct signature of entanglement. {Yet, this is not the case for the class of \DBR{non}-passively-separable states, where those changes of basis do not disentangle the state}. Because this entanglement is generated by non-Gaussian operations and hidden in the non-Gaussian features of the state, we will here refer to it as \textit{non-Gaussian entanglement} \cite{Namiki2012, Walschaers2021}. The goal of this work is to find a practical way to detect this \DBR{non-Gaussian entanglement.}


In order to characterize non-Gaussian entanglement\DBR{,} a number of criteria based on high-order moments and on uncertainty relations of different classes of operators have been proposed \cite{Shchukin2005, Miranowicz2006, Shchukin2016, Agarwal2005, Hillery2006}. Nevertheless, these criteria are far from being feasible with current experimental methods. Other more experimentally-friendly criteria are based on the Shannon entropy and the fidelity of teleportation in quantum channels \cite{Walborn2009, Nha2012}. Here we tackle the problem from an operational point of view: non-Gaussian quantum correlations are also known to improve metrological sensitivity, the performance of quantum key distribution and quantum teleportation protocols \cite{Strobel2014, Hu2020, Opartny2000,Filippov2014}. The advantage of relying on the improvement of quantum protocols is that the detected entanglement is useful by design. In this \DBR{article}, we will focus specifically on metrological protocols, where quantum estimation tools have been devised to witness entanglement \cite{Pezze2009, Gessner2016, Gessner2017}. These witnesses are based on the fact that metrological sensitivity determines precision of measurements and this sensitivity is limited for separable states. This can be used to detect entanglement. Two powerful assets of these sensitivity-based witnesses are i) they do not make assumptions about the quantum state --Gaussianity, purity, etc., and ii) they contain information about all high-order moments. 

By using the approach of refs. \cite{Gessner2016, Gessner2017} taking into account the limitations of CV quantum optics, we propose a general protocol based solely on homodyne detection, using both the variance of the measurement outcomes and the joint measurement statistics. Our protocol is efficient in terms of resources as the parameter estimation is done in postprocessing using solely the data collected by homodyne detection. We show its relevance analyzing an important class of non-Gaussian entangled states: photon-subtracted states. We demonstrate that our protocol is pertinent for the detection of non-Gaussian entanglement and that it is experimentally feasible.

The article is organized as follows: We first present our protocol to detect entanglement through homodyne detection and postprocessing of the joint probability distribution based on the metrological witness introduced in \cite{Gessner2016, Gessner2017} in Section~\ref{sec:II}. We then present in Section~\ref{sec:III} the probe states \DBR{that} we will use to test our non-Gaussian entanglement witness. In Section~\ref{sec:IV} we analyze which parameter is best suited to measure entanglement in our metrological protocol and calculate entanglement in an ideal case. In Section~\ref{sec:V}  we study a realistic case taking into account unbalanced input squeezing, losses and discretization of the measurement outcomes. We finally discuss possible experimental implementations of our scheme, their limitations and feasibility in Section \ref{sec:VI} and we present our conclusions in Section \ref{sec:VII}.


\section{Entanglement detection via local homodyne detection and postprocessing}\label{sec:II}

We consider here the following problem: two experimenters, Alice and Bob, who share an optical quantum state $\hat{\rho}_{A B}$, want to elucidate if their shared state is entangled or not, while minimizing the amount of experimental resources. If the input state is Gaussian, they just need to measure the variances of linear combinations of optical-field quadratures and apply second-order moment-based criteria like for instance those of Duan {\it et al.}, Simon or Giovanetti {\it et al.} \cite{Duan2000, Simon2000, Giovanetti2003}. This can be easily implemented experimentally using homodyne detection. However, the larger class of non-Gaussian states do not always present entanglement that can be revealed by second-order moment-based criteria. Particularly, the majority of entanglement criteria for quantum states with purely non-Gaussian correlations are based on either carrying out full quantum-state tomography \cite{Ourjoumtsev2009} or measuring high-order moment correlations, protocols which are very demanding experimentally. 

\begin{figure}[t]
\includegraphics[width=0.5\textwidth]{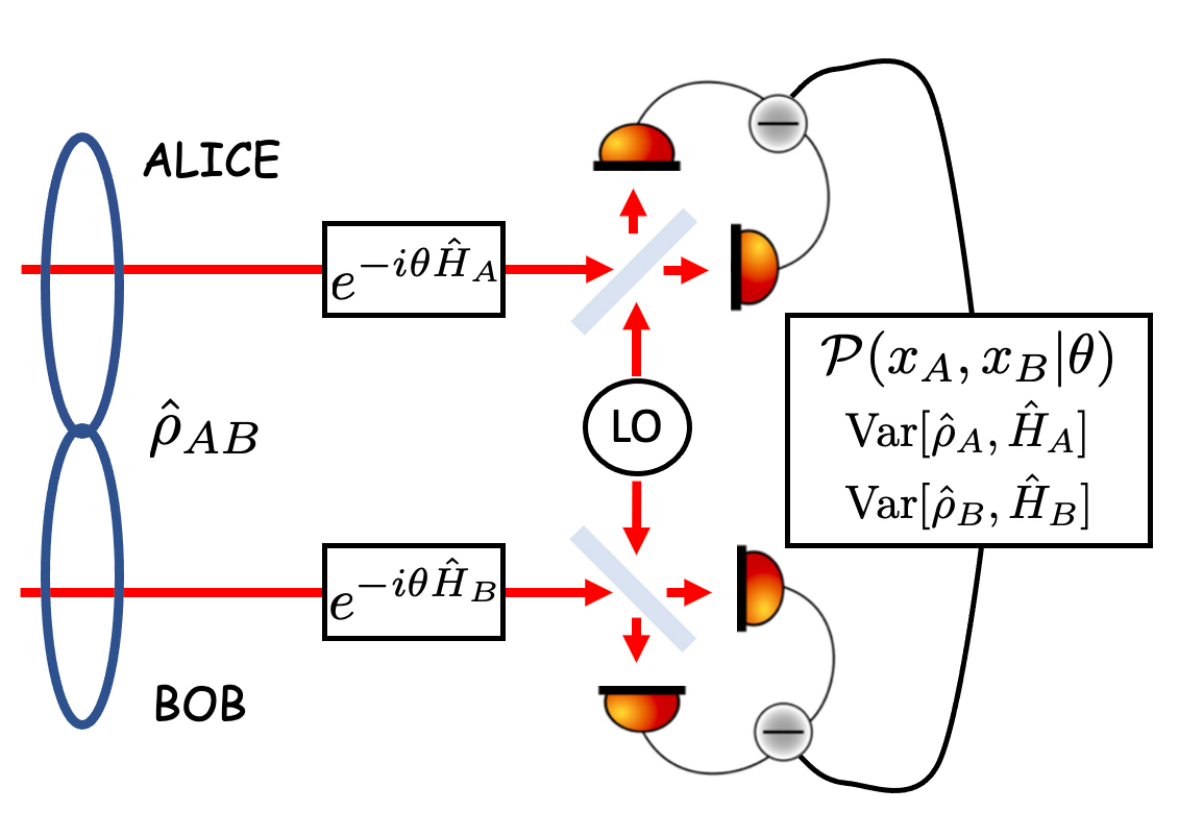} 
\hspace{0cm}\caption{\label{F0}\small{Sketch of the proposed metrological protocol for entanglement detection. Alice and Bob share a quantum state $\hat{\rho}_{A B}$. They jointly estimate a parameter $\theta$ generated by two local Hamiltonians $\hat{H}_{A/B}$. Using two homodyne detectors with a common phase reference, Alice and Bob can retrieve the parameter-dependent joint probability distribution $\mathcal{P}(x_{A}, x_{B}|\theta)$, and thus the Fisher information related to this parameter estimation, and the local variances of the Hamiltonians $\hat{H}_{A/B}$. With this in hand, Alice and Bob can jointly compute the metrological entanglement witness of Equation (\ref{Ent1}).}}
\end{figure}

Here, we apply a metrological protocol to detect entanglement. Alice and Bob share information in order to estimate jointly a parameter $\theta$ generated by a Hamiltonian $\hat{H}=\hat{H}_{A} + \hat{H}_{B}$ that acts locally on both subsystems such that$\hat{\rho}_{A B}^{\theta}=e^{-i\theta \hat{H}} \hat{\rho}_{A B} e^{i\theta \hat{H}}$ (see Figure \ref{F0}). \MII{We define the joint probability density, i.e., the conditional probability to obtain a set of local measurement outcomes $(\xi_A,\xi_B)$ given the parameter $\theta$, as 
\begin{equation*}
   \mathcal{P}(\xi_A, \xi_B|\theta) =\text{Tr}[\hat{\rho}_{A B}^{\theta} \hat{\Pi}],
\end{equation*} 
where $\hat{\Pi}$ is a positive-operator valued measure (POVM) such that $\int \hat{\Pi} \, d^2\boldsymbol{\xi} = \mathds{1}$, where $d^2\boldsymbol{\xi} = d\xi_A \, d\xi_B$. The metrological protocol consists of measuring the Fisher information (FI) defined as 
\begin{equation}\label{FIC}
F(\mathcal{P}(\xi_A, \xi_B|\theta))=\int_{\mathbb{R}^{2}} \mathcal{P}(\xi_A, \xi_B|\theta) \left(\frac{\partial \mathcal{L}(\xi_A, \xi_B|\theta)}{\partial \theta}\right)^2 d^2\boldsymbol{\xi},
\end{equation}
where  $\mathcal{L}(\xi_A, \xi_B|\theta)=\log(\mathcal{P}(\xi_A, \xi_B|\theta))$ represents the logarithmic likelihood related to the probability density $\mathcal{P}(\xi_A, \xi_B|\theta)$. }
In our case, as illustrated in Figure \ref{F0}, the observables will correspond to local homodyne measurements $\hat{\xi}_A = \cos \phi_A \hat{x}_A + \sin \phi_A \hat{p}_A$ and $\hat{\xi}_B= \cos \phi_B \hat{x}_B + \sin \phi_B \hat{p}_B$, where $\phi_A$, $\phi_B$ are two angles, and $\hat{x}_A, \hat{x}_B, \hat{p}_A , \hat{p}_B$ are the quadratures operators defined from the annihilation operators as
\begin{equation}
    \hat{a}_I = \frac{\hat{x}_I + i \hat{p}_I}{2}, \quad I \in \{ A,B\}.
\end{equation}
The quadrature operators thus satisfy the commutation rules $[\hat{x}_I, \, \hat{p}_J] = 2 i \delta_{IJ}, \,  I,J \in \{ A,B\}$.

Then, if $\hat{\rho}_{A B}$ is separable, the FI of Equation \eqref{FIC} for a state $\hat{\rho}_{A B}^{\theta}$ generated by $\hat{H}$ is upper bounded by \cite{Gessner2016, Gessner2017}
\begin{equation}\label{FQ}
F(\mathcal{P}(\xi_A, \xi_B|\theta)) \leq 4 \text{Var}[\hat{\rho}_A,\hat{H}_A]+4 \text{Var}[\hat{\rho}_B,\hat{H}_B], 
\end{equation}
where $ \hat{\rho}_{A/B}$ are the reduced density matrices for systems $A$ and $B$, respectively. Because this is a necessary condition for separability, its violation is a sufficient criterion for entanglement. 

Therefore, we can introduce the following metrological witness of entanglement
\begin{equation}\label{Ent1}
\begin{split}
E=F(\mathcal{P}(\xi_A, \xi_B|\theta)) - 4 (\text{Var}[\hat{\rho}_A,\hat{H}_A]+ \text{Var}[\hat{\rho}_B,\hat{H}_B])>0. 
\end{split}
\end{equation}
This inequality can reveal entanglement but not its origin --Gaussian or non-Gaussian.  \MI{However, as we will discuss below, this witness can reveal entanglement generated by non-Gaussian operations where Gaussian criteria fail to detect it. Therefore, this \textit{non-Gaussian entanglement} leads to a better sensitivity to estimate a given parameter $\theta$ than one would have obtained with a separable state and becomes a useful resource for metrology.} 


The \MI{other} interest of the witness \eqref{Ent1} is that it also holds for any state, pure or mixed, and its major asset is the practicability of its computation. Homodyne measurements in each mode with a common phase reference allow us to access experimentally i) the joint probability distribution $\mathcal{P}(\xi_{A}, \xi_{B}|\theta)$, and thus the FI, and ii) the variances associated to the local generators, enabling to test the entanglement witness given by Equation (\ref{Ent1}). Moreover, in some cases (see Section \ref{sec:IV}) the parameter-dependence of the joint probability distribution $\mathcal{P}(\xi_{A}, \xi_{B}|\theta)$ can be generated in postprocessing applying appropriate transformations directly to the joint probability distribution as $\mathcal{P}(\xi_{A}, \xi_{B}|\theta)=\mathcal{P}(U_{\theta}(\xi_{A}),U_{\theta}(\xi_{B}))$, with $U_{\theta}(\xi_{A/B})$ the transformation related to the Hamiltonian $\hat{H}_{A/B}$ in the quadrature space $\xi_{A/B}$ \cite{Nieto1997,Comment1}. This important feature avoids to apply impractical inline transformations to the state simplifying greatly the detection of entanglement.

\MII{In Equation \eqref{Ent1}, the FI is always bounded by the Quantum Fisher information (QFI) defined as 
\begin{equation}
    F_Q[\hat{\rho}_{A B},\hat{H}] = \underset{\hat{\Pi}}{\text{max}} \, F(\text{Tr}[\hat{\rho}_{A B}^{\theta} \hat{\Pi}]).
\end{equation}
This bound is reached when the measurement observable is optimized over all possible POVM $\hat \Pi$ \cite{Braunstein1994}. In this case, 
the entanglement witness \eqref{Ent1} is maximized and we denote this quantity by $E_Q$:
\begin{equation}
\begin{split}
E_Q & \equiv \max_{\hat{\Pi}}{E} \\
&=  F_Q[\hat{\rho}_{A B},\hat{H}] - 4 (\text{Var}[\hat{\rho}_A,\hat{H}_A]+ \text{Var}[\hat{\rho}_B,\hat{H}_B]).
\end{split}
\end{equation}
Note that $E_Q$ might not be attained when we restrict ourselves to homodyne measurements since it might not correspond to the optimal POVM to saturate the QFI.
}

For pure states we can easily obtain the QFI from the variance of the generator $\hat{H}$ of the parameter $\theta$ as  
\begin{equation}\nonumber
F_Q[\rho_{A B}, \hat{H}]=4 \text{Var}[\hat{\rho}_{AB}, \hat{H}].
\end{equation}
Applying this identity into Equation (\ref{Ent1}) we obtain the following simple condition for entanglement
\begin{equation}\label{Ent}
E_Q 
=8\, \text{Cov}[\rho_{A B}; \hat{H}_{A}, \hat{H}_{B}] > 0,
\end{equation}
\MII{where $\text{Cov}[\rho_{A B}; \hat{H}_{A}, \hat{H}_{B}] = \langle \hat{H}_{A} \hat{H}_{B}\rangle - \langle \hat{H}_{A} \rangle \, \langle \hat{H}_{B}\rangle$. As a consequence any correlation that is seen in a bipartite pure state is a signature of entanglement.}

\section{Application to photon-subtracted states}\label{sec:III}

The protocol described in the previous section is valid for any CV system, regardless of the nature of the state under consideration, as long as one has access to the probability distributions of each subsystem. In this section we introduce the states that we will use as a probe of our non-Gaussian entanglement criterion, namely, photon-subtracted states. In particular we will analyze bipartite states without Gaussian correlations in order to focus on their non-Gaussian features.

We consider two-mode photon subtracted states. This class of states has been demonstrated in optical systems using different degrees of freedom, such as polarization or frequency modes \cite{Ourjoumtsev2009, Ra2020}. In Section \ref{sec:VI} we will explain in detail different experimental methods for their production. Let us consider two independent single-mode squeezed states respectively related to Alice and Bob
\MII{\begin{equation}\label{Sq}
\vert \Psi_{0}\rangle =\hat{S}_{A}(r_{A}
) \hat{S}_{B}(r_{B}
) \vert 0 0 \rangle,
\end{equation}
where $\hat{S}_{I}(r_{I})=\text{exp}\{(-r_{I}/2)(\hat{a}_{I}^2 -\hat{a}_{I}^{\dagger 2} )\}$ is the single-mode squeezing operator, and $r_{I} \in \mathbb{R}$ is the squeezing parameter for each mode $I=A,B$. The amount of squeezing in decibels is given by $s_{I}=10 \log_{10} (e^{-2 r_{I}})$.

In what follows we analyze two cases: in-phase squeezing ($r_A>0$ and $r_B>0$) and in-quadrature squeezing ($r_A>0$ and $r_B<0$).
} 
Equation \eqref{Sq} corresponds to a Gaussian state and all its information is encoded in the covariance matrix $V_{0}=\text{diag}(e^{-2 r_{A}}, e^{2 r_{A}}, e^{-2 r_{B}},e^{2 r_{B}})$, written with respect to the vector of amplitude and phase quadratures in each mode $\vec{\xi}=(x_{A},p_{A},x_{B},p_{B})^{T}$. Note that $V_{0}$ does not present off-diagonal terms, thus the input Gaussian state is fully separable. 

\begin{figure}[t]
\includegraphics[width=0.45\textwidth]{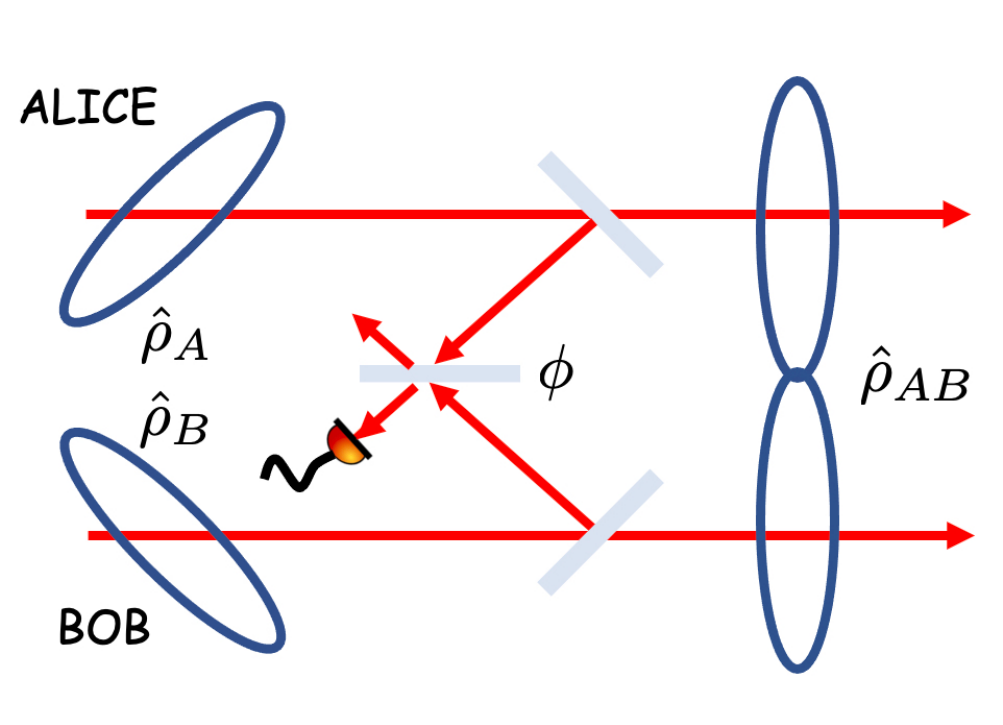} 
\hspace{0cm}\caption{\label{FigX}\small{Sketch of an optical setup for delocalized single-photon subtraction. Alice and Bob prepare two
squeezed states in given optical modes. A small fraction of each mode power is diverted to a common beam splitter with a transmittivity controlled by a parameter $\phi$. An event measured by a single-photon detector heralds the subtraction of a photon delocalized between the two modes.
}}
\end{figure}

\begin{figure}[t]
  \centering
    \subfigure{\includegraphics[width=0.44\textwidth]{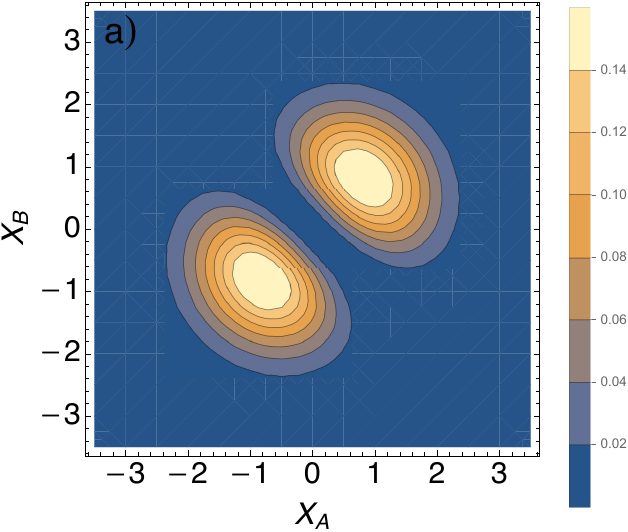}}
    \subfigure{\includegraphics[width=0.44\textwidth]{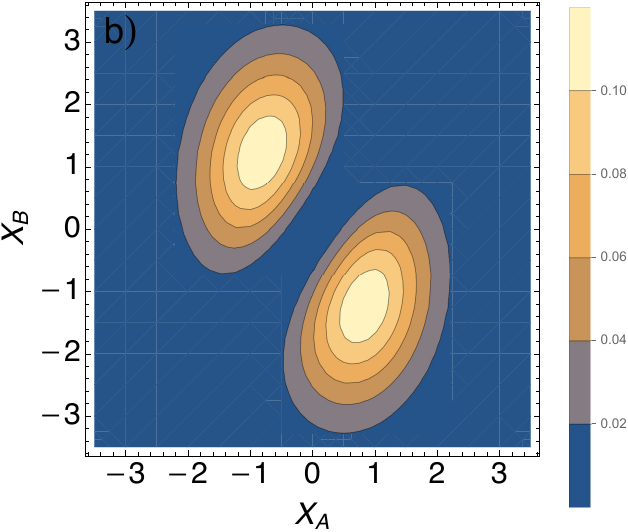}}
\hspace{0cm}\caption{\label{F1}\small{Contour plots of the joint probability distribution for a single-photon subtracted state given by Equation (\ref{UnSymPsi}) with $\phi=\pi/4$ and a) $r_{A}=r_{B}=0.2$, b) $r_{A}=-r_{B}=0.2$ ($\vert s_{A/B}\vert = 1.74$ dB).}}
\end{figure}

Next, we perform a delocalized subtraction of $n$ photons in the same or different modes on this state. This operation produces in general a superposition of squeezed Fock states. For multiphoton subtraction \DBR{the resulting state is} 
\DB{
\begin{equation}\nonumber
\vert\Psi\rangle \propto \prod_{j=1}^{n}(\cos(\phi_{j}) \hat{a}_{A}+\sin(\phi_{j}) \hat{a}_{B}) \vert \Psi_{0}\rangle,
\end{equation}
where the parameters $\phi_{j}$ control the probability of subtraction in each mode for each subtraction $j=1, \dots, n$.
For $n=1$ we have \DBR{\cite{Agarwal2013}}
\begin{align}\nonumber
&\vert\Psi\rangle \propto (\cos(\phi) \hat{a}_{A}+\sin(\phi) \hat{a}_{B}) \vert \Psi_{0}\rangle = \\ \nonumber
&\hat{S}_{A}(r_{A}) \hat{S}_{B}(r_{B}) (\cos(\phi) \sinh(r_{A})\vert 1 0 \rangle + \sin(\phi) \sinh(r_{B})\vert 0 1 \rangle),
\end{align}
\DBR{where we have used the Bogolyubov transformation} $\hat{S}_{I}^{\dag}(r_{I})\hat{a}_{I}\hat{S}^{}_{I}(r_{I})=\cosh(r_{I})\hat{a}_{I}+\sinh(r_{I}) \hat{a}^{\dag}_{I}$. A sketch of this operation is shown in Figure \ref{FigX}. The wavefunction} of a single-photon subtracted state in the amplitude quadratures of the optical field is given by
\begin{align} \label{UnSymPsi}\nonumber
\Psi(x_{A},&x_{B})\equiv \langle x_{A}, x_{B} \vert\Psi\rangle \propto  e^{ -\frac{e^{2 r_{\scriptscriptstyle A}} x_{A}^{2} + e^{2 r_{ \scriptscriptstyle B}} x_{B}^{2}}{4} } \\
&\times[(e^{2 r_{A}}-1) \cos{(\phi)} x_{A} + (e^{2 r_{B}}-1) \sin{(\phi)} x_{B}].
\end{align}
Examples of joint probability distributions $\mathcal{P}(x_{A},x_{B})=\vert \Psi(x_{A},x_{B}) \vert^{2}$ for a photon subtracted state given by Equation (\ref{UnSymPsi}) with $\phi=\pi/4$ and $r_{A}=r_{B}=0.2$, $r_{A}=-r_{B}=0.2$, are respectively shown in Figure \ref{F1} a) and b).

\DB{
For $n=2$ we can consider for instance subtraction in the same \DBR{($\phi_{1}=\phi_{2}=\phi$)} or orthogonal modes \DBR{($\phi_{1}=\phi$, $\phi_{2}=\phi+\pi/2$)} as
\begin{align}\nonumber
\vert\Psi_{\parallel}\rangle &\propto (\cos(\phi) \hat{a}_{A}+\sin(\phi) \hat{a}_{B})^2 \vert \Psi_{0}\rangle,\\ \nonumber
\vert\Psi_{\perp}\rangle &\propto (\cos(\phi) \hat{a}_{A} + \sin(\phi) \hat{a}_{B}) \, (\sin(\phi) \hat{a}_{A} - \cos(\phi) \hat{a}_{B}) \vert \Psi_{0}\rangle.
\end{align}
Taking in-phase squeezing, $r_{A}=r_{B}\equiv r$ and subtraction along $\phi=\pi/4$, we obtain
\begin{align} \label{eq:psi_parallel}
\vert\Psi_{\parallel}\rangle \propto  \hat{S}_{A}(r) \hat{S}_{B}(r) &(\cosh(r) \vert 00 \rangle + \sinh(r) \vert 11 \rangle \\ \nonumber
+&(1/\sqrt{2})(\vert 20 \rangle + \vert 02 \rangle)),
\\ \nonumber
\vert\Psi_{\perp}\rangle \propto\hat{S}_{A}(r) \hat{S}_{B}(r)&(\vert 20 \rangle - \vert 02 \rangle).
\end{align}
Thus, subtracting photons along different axes produce states with completely different entanglement features. Examples of joint probability distributions for two-photon subtracted states are gathered in the Supplementary Material. 
}

\MII{Coming back to the case of single-photon subtracted states,} the entanglement present in these states is not grasped by Gaussian entanglement witnesses: this can be generally understood from the covariance matrix of a photon-subtracted state. Ref. \cite{Walschaers2017} shows that this covariance matrix for a single-photon subtraction can generally be written as 
\begin{equation} \label{Vnoisy}
    V = V_0 + 2 \frac{(V_0 - \mathds{1})P(V_0 - \mathds{1})}{{\rm Tr}[(V_0 - \mathds{1})P]},
\end{equation}
where $V_0$ is the initial Gaussian state's covariance matrix and $P$ is a matrix that projects on the phase space axes associated with the mode of photon subtracted states. In our present case, we find that
\begin{equation} \nonumber \begin{split}
    P = \begin{pmatrix}
 \cos ^2(\phi ) & 0 & \frac{1}{2}\sin (2 \phi ) & 0 \\
 0 & \cos ^2(\phi ) & 0 & \frac{1}{2}\sin (2 \phi ) \\
\frac{1}{2}\sin (2 \phi ) & 0 & \sin ^2(\phi ) & 0 \\
 0 & \frac{1}{2}\sin (2 \phi ) & 0 & \sin ^2(\phi ) \\
\end{pmatrix}.
\end{split}
\end{equation}
\MII{ Since $P$ is a positive matrix and $V_0 - \mathds{1}$ a symmetric matrix, $(V_0 - \mathds{1})P(V_0 - \mathds{1})$ is also positive. Thus, as we see from Equation \eqref{Vnoisy}, on the level of the covariance matrix the photon subtraction only adds Gaussian noise.} This implies that no additional entanglement can be witnessed by purely looking at the covariance matrix \cite{Hyllus2009}. As a consequence, since we set $V_{0}=\text{diag}(e^{-2 r_{A}}, e^{2 r_{A}}, e^{-2 r_{B}},e^{2 r_{B}})$, we find that $V$ should not display any entanglement. 

\MI{On the contrary, for two-photon subtracted states, Gaussian witnesses can reveal entanglement in some cases.
This results from constructive interferences between both photon subtractions which induce quantum correlations at the level of the covariance matrix. In particular, when the two subtractions happen in the same mode along $\phi= \pi/4$ and when the squeezing is low, we can see directly from expression \eqref{eq:psi_parallel} that Gaussian correlations appear. Indeed, in that case, a part of $\vert\Psi_{\parallel}\rangle $ is proportional to $|00\rangle + \tanh(r) |11\rangle$, which is the expansion of a two-mode squeezed state at order \DBR{$O(r)$}, and thus induces Gaussian correlations that can be witnessed at the level of the covariance matrix.

However, Gaussian witnesses are very sensitive to the parameters of the subtracted states: for instance, when the squeezing parameters of the initial Gaussian state are too high or when the subtractions happen in orthogonal modes, Gaussian witnesses cease to work. The metrological protocol proposed in this paper circumvents this issue since it hinges on the Fisher information which can reveal correlations overlooked by witnesses only based on second order moments. As discussed in the following sections, the metrological criterion will always be able to detect entanglement for two-photon subtracted states regardless to their properties. We discuss in particular at the end of section V an interesting case where Gaussian witnesses fail to detect entanglement while the metrological criterion works and can be used directly with a limited amount of homodyne data.}

\section{Ideal detection of non-Gaussian entanglement}\label{sec:IV}

In order to decide which Hamiltonian $\hat{H}$ is best suited to witness entanglement we can calculate theoretically $E_Q$ through Equation \eqref{Ent}. This can guide us deciding which parameter is best suited to detect entanglement of a given quantum state in a realistic scenario. Below, we use the estimation of parameters related to the four single-mode Gaussian gates in CV quantum optics. Namely: displacement, phase-shift, shearing and squeezing. We analyze in which cases the joint estimation of these parameters reveals the entanglement of \DBR{the single-photon subtracted state} given by Equation (\ref{UnSymPsi}). For the sake of simplicity, we focus here on the case $\phi=\pi/4$. A generalization to any angle $\phi$ and an equivalent analysis for two photon-subtracted states can be found in the Supplemental Material.

\subsection{Displacement} \label{subsec:d}

A displacement of $\theta$ along the axis $x_{A}=\pm x_{B}$ is produced by the following operator
\begin{equation}\nonumber
\hat{D}(\theta)=e^{-i \theta (\hat{p}_{A} \pm \hat{p}_{B})/2 }.
\end{equation}
The Hamiltonian related to this displacement operator is $H_{\pm}= (\hat{p}_{A} \pm \hat{p}_{B})/2$. The optimal entanglement witness $E_{Q}$ obtained with displacement operators along the axis $x_{A}=\pm x_{B}$ for a pure photon-subtracted state given by Equation (\ref{UnSymPsi}) 
with $\phi=\pi/4$ 
is
\begin{equation}\label{Ed}
E_{Q}=\pm 2 e^{r_A + r_B} \cos(\epsilon),
\end{equation}
with 
\begin{equation}\nonumber
\cos(\epsilon)=\frac{2\sinh(r_A) \sinh(r_B)}{\sinh^{2}(r_A) + \sinh^{2}(r_B)}.
\end{equation}

\begin{figure}[t]
  \centering
    \subfigure{\includegraphics[width=0.42\textwidth]{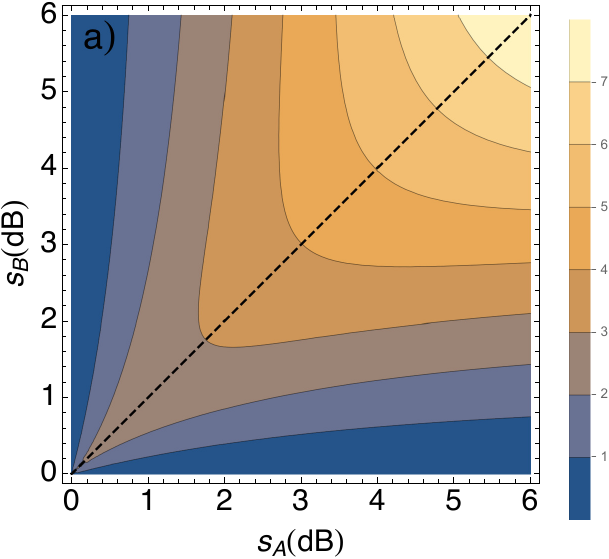}}
    \subfigure{\includegraphics[width=0.45\textwidth]{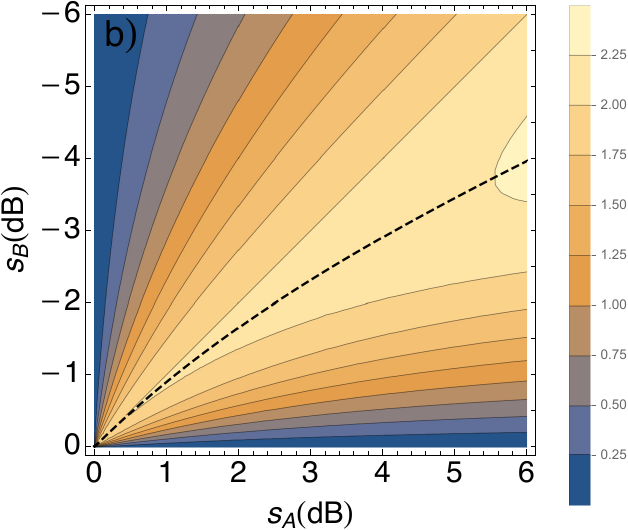}}
\vspace {0cm}\,
\hspace{0cm}\caption{\label{F3}\small{\MII{Displacement-estimation entanglement witness $E_{Q}$ for one-photon subtracted states \eqref{UnSymPsi}} given by Equation (\ref{Ed}) versus squeezing of the input squeezed states $s_A$ and $s_B$. a) Displacement along $x_{A}=x_{B}$ optimizes $E_{Q}$ for states with in-phase input squeezing. b) Displacement along $x_{A}=- x_{B}$ optimizes $E_{Q}$ for states with in-quadrature input squeezing. The black dashed lines correspond to the maximum value of $E_{Q}$.}}
\end{figure}

Displacement along either $x_{A}=x_{B}$ or $x_{A}=-x_{B}$ detects entanglement respectively for in-phase squeezing ($r_A, r_B >0$) and in-quadrature --orthogonal-- squeezing ($r_A>0$, $r_B <0$). Figures \ref{F3}a and \ref{F3}b show contour plots of optimal entanglement witness $E_{Q}$ in the two cases. States with in-phase input squeezing show always a larger \MII{violation of $E_Q$} due to the argument $r_A + r_{B}$ in Equation \eqref{Ed}. \MII{One can also optimize the input squeezing parameters $r_A$ and $r_B$ to maximize the witness $E_Q$. These values correspond to the black dashed lines in Figures \ref{F3}a and \ref{F3}b. For in-phase squeezing the witness reaches its maximum at $r_A=r_{B}$ and is given by $E_{Q}=2 e^{2 r_{A}}$. Likewise, for in-quadrature squeezing the maximum value of $E_{Q}$ is not along the diagonal (see the dashed line in Figure \ref{F3}b), but below it. For a given value of $r_{A}$, $E_{Q}$ reaches its maximum for $r_{B}=\log{(1/(1+2\sinh{(r_{A})})^{1/2})}$ and is given by}
\begin{equation}\nonumber
E_{Q}=\frac{2 \,e^{r_{A}}}{1+\sinh{(r_{A})}}.
\end{equation}

The shapes of Figures \ref{F3}a and \ref{F3}b can be explained in terms of the symmetries of the two functions that compose Equation \eqref{Ed}: $\pm \cos(\epsilon)$ is a symmetric function with respect to the diagonal $s_{A}=s_{B}$ for every input squeezing, whereas $2 e^{r_A + r_B}$ is symmetric with respect to the diagonal (antidiagonal, in this case along $s_{A}-s_{B}=6$ dB) for in-phase (in-quadrature) squeezing.

Importantly, we obtain the same result calculating the entanglement witness through Equation (\ref{Ent1}), $E=E_{Q}$, indicating that the FI saturates the QFI. The result of Equation \eqref{Ed} is particularly interesting because, following Equation (\ref{Ent}), second order moments of the distribution reveal entanglement with a non-Gaussian origin.

Recently, M. Tian {\it et al.} analyzed the multipartite entanglement in a nondegenerate triple photon state using a metrological criterion \cite{Tian2022}. They claimed there that non-Gaussian entanglement cannot be sufficiently captured by linear quadratures, i.e. displacements. While this is the case for triple photon states, we have shown that it does not hold in general: displacements can detect non-Gaussian entanglement of photon-subtracted states.

\subsection{Phase shift} \label{subsec:ps}
\begin{figure}[t]
  \centering
    \subfigure{\includegraphics[width=0.42\textwidth]{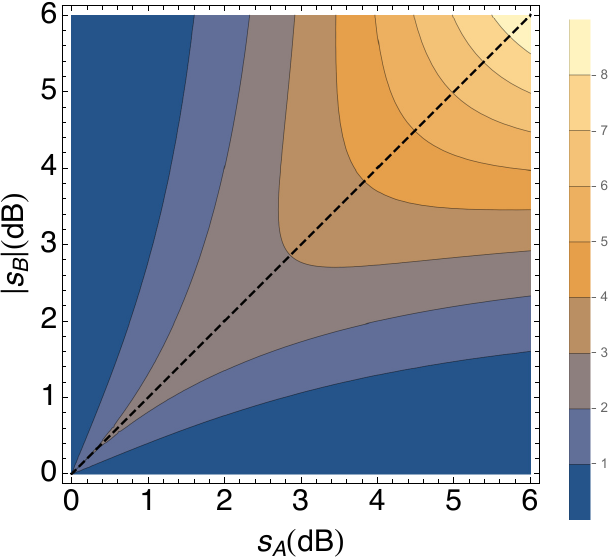}}
\vspace {0cm}\,
\hspace{0cm}\caption{\label{F3a}\small{Phase-estimation entanglement witness $E_{Q}$ for one-photon subtracted states \eqref{UnSymPsi} given by Equation \eqref{EdPhase} versus squeezing of the input squeezed states $s_A$ and $|s_B|$. The black dashed line corresponds to the maximum value of $E_{Q}$.}}
\end{figure}

The phase-shift operator 
\begin{equation}\nonumber
\hat{R}(\theta)=e^{-i \theta (\hat{N}_{A} \pm \hat{N}_{B}) } \propto e^{-i \theta (\hat{x}_{A}^2 +\hat{p}_{A}^2 \pm \hat{x}_{B}^2 \pm \hat{p}_{B}^2)/4}
\end{equation}
rotates the state by a phase $\theta$ in local phase subspaces in either clockwise-clockwise ($+$) or clockwise-counterclockwise ($-$) directions. The related Hamiltonian is $H_{\pm}= \hat{N}_{A} \pm \hat{N}_{B}$. The optimal entanglement witness is in this case 
\begin{equation}\label{EdPhase}
E_{Q}=\mp 2 \cosh(2 r_{A})\cosh(2 r_{B}) \cos^{2}(\epsilon).
\end{equation}
Entanglement is always detected for clockwise-counterclockwise ($-$) phase shifts, but not for clockwise-clockwise ($+$) as it is just a global phase shift. Figure \ref{F3a} shows contour plots of the entanglement witness $E_{Q}$ for different values of squeezing. Notably, the detection of entanglement does not depend on the sign of the input squeezed states as $E_{Q}$ is invariant under change of sign of the squeezing parameters $r_{A/B}$. The detected entanglement witness is maximum for $r_A=|r_{B}|$ (dashed line along the diagonal in Figure \ref{F3a}) being $E_{Q}=2 \cosh^{2}(2 r_{A})$.

One can wonder if the Fisher information in Equation \eqref{Ent1} reaches the QFI in this case. While measuring the joint probability distribution in the $(x_1,x_2)-$plane was enough to obtain the maximal value of the FI and saturates the QFI for the displacement estimation, here the situation is a bit more complicated.
For simplicity, we consider the case $r_A=r_{B}$ in what follows. The FI can be optimized by finding the set of angles ($\phi_A, \phi_B$) of the measurement outcomes $\xi_A = \cos \phi_A x_A - \sin \phi_A p_A$, $\xi_B = \cos \phi_B x_B - \sin \phi_B p_B$ for which the joint probability distribution $P(\xi_A, \xi_B|\theta)$ leads to the best value of the FI (see the Supplemental Material).
However, we find that such local rotations are not enough to saturate the QFI, and that only a mixing of modes $A$ and $B$ before the homodyne detectors can lead to a saturation of the QFI. It is indeed possible to prove that a non-local rotation of $-\pi/4$ between modes $A$ and $B$ and measuring the joint probability distribution $\mathcal{P}(x^{\prime}_A,p^{\prime}_B|\theta)$, with $x^{\prime}_A = (x_A - x_B)/\sqrt2$ and $p^{\prime}_B = (p_A + p_B)/\sqrt2$ is needed to saturate the QFI.

\subsection{Shearing} \label{subsec:sh}
The shearing --also known as phase-gate-- operator 
\begin{equation}\nonumber
\hat{\mathcal{S}}(\theta)=e^{-i \theta (\hat{x}_{A}^2 \pm \hat{x}_{B}^{2})/4 }
\end{equation}
shears the state with respect to the axes $x_{A}$ and $\pm x_{B}$ by a gradient of $\theta$. The related Hamiltonian is $H_{\pm}= (\hat{x}_{A}^2 \pm \hat{x}_{B}^{2})/4$. The optimal entanglement witness is in this case 
\begin{equation}\label{EdShear}
E_{Q}=\mp \frac{e^{-2(r_A + r_B)}}{2} \cos^{2}(\epsilon).
\end{equation}
Thus, shearing with respect to $x_{A}$ and $x_{B}$ does not detect entanglement. However, shearing with respect to $x_{A}$ and $-x_{B}$ captures it. Note that in this case the entanglement witness $E_Q$ is maximized for $r_{A/B}<0$, i.e. squeezing along the quadratures $p_{A/B}$, unlike displacement and phase estimation where $E_{Q}$ is maximized for squeezing along $x_{A/B}$.

\begin{figure}[t]
  \centering
    \subfigure{\includegraphics[width=0.45\textwidth]{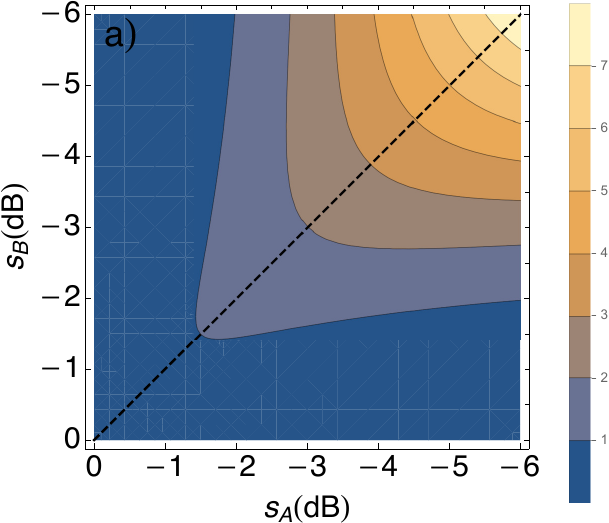}}
    \subfigure{\includegraphics[width=0.45\textwidth]{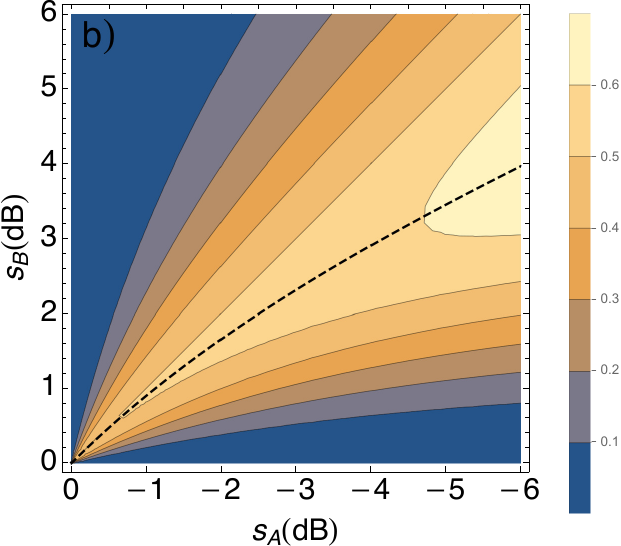}}
\vspace {0cm}\,
\hspace{0cm}\caption{\label{F3b}\small{Shearing-estimation entanglement witness $E_{Q}$ for one-photon subtracted states \eqref{UnSymPsi} given by Equation \eqref{EdShear} versus squeezing of the input squeezed states $s_A$ and $s_B$. a) $E_{Q}$ for states with input squeezing along the same quadrature. b) $E_{Q}$ for states with input squeezing along different quadratures. The black dashed lines correspond to the maximum value of $E_{Q}$.}}
\end{figure}

Figures \ref{F3b}a and \ref{F3b}b show contour plots of $E_{Q}$ in the cases of input squeezing along the same quadrature (a) or along different quadratures (b). For input squeezing along the same quadratures the detected entanglement witness is maximum again for $r_A=r_{B}$ (dashed line along the diagonal in Figure \ref{F3b}a) and given by $E_{Q}=e^{-4 r_{A}}/2$. However, for input squeezing along different quadratures, the maximum value for $E_Q$ is below the diagonal, as it also happens for displacement. For a given value of $r_{A}$, the maximum $E_{Q}$ is obtained for $r_{B}=(-r_{A}+\log(1+e^{r_{A}}-e^{2r_{A}}))/2$ (dashed line in Figure \ref{F3b}b) and is given by
\begin{equation}\nonumber
E_{Q}=\frac{e^{-2r_{A}}}{2(-1+\sinh(r_{A}))^{2}}.
\end{equation}
The shape of Figures \ref{F3b}a and \ref{F3b}b is explained in the same way as for displacement. 

Here again, we optimize the FI to see if it is possible to reach the bound $E=E_Q$. The same analysis as in the case of the phase-shift operator (by performing local rotations before the homodyne detection) is summarized in the Supplemental Material. The same conclusion follows: the FI never reaches the QFI, and only a non-local rotation of -$\pi/4$ between modes $A$ and $B$ leads to a saturation of the QFI \DBR{from quadrature measurements}.

\subsection{Squeezing} \label{sec:subsecIVD}
The squeezing operator 
\begin{equation}\nonumber
\hat{S}(\theta)=e^{-i \theta (\hat{x}_{A}\hat{p}_{A}+\hat{p}_{A}\hat{x}_{A} \pm \hat{x}_{B}\hat{p}_{B} \pm \hat{p}_{B}\hat{x}_{B})/4 }
\end{equation}
squeezes the position quadratures of modes A and B by a factor of $e^{\theta}$ ($+$) or squeezes the position quadratures of A by $e^{\theta}$ and stretches those of B by $e^{-\theta}$ ($-$). The related Hamiltonian is $H_{\pm}= (\hat{x}_{A}\hat{p}_{A}+\hat{p}_{A}\hat{x}_{A} \pm \hat{x}_{B}\hat{p}_{B} \pm \hat{p}_{B}\hat{x}_{B})/4$. The optimal entanglement witness is here
\begin{equation}\nonumber
E_{Q}=0.
\end{equation}
Interestingly, the joint estimation of the squeezing parameter does not detect entanglement in any of the above two cases. 

\subsection{Comparison and resource evaluation}

\begin{figure}[t]
\includegraphics[width=0.47\textwidth]{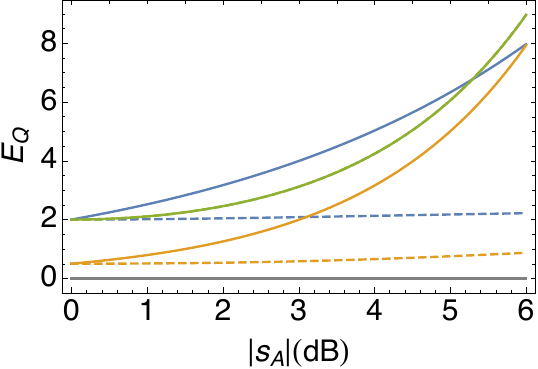} 
\hspace{0cm}\caption{\label{F2}\small{\MII{Maximal value of the entanglement witness $E_{Q}$, i.e., when input squeezing parameters $r_A$ and $r_B$ are optimized to follow the dashed curves in Figures \ref{F3}, \ref{F3a}, and \ref{F3b}, versus squeezing in Alice's mode}: displacement estimation (blue), shearing estimation (orange), phase shift estimation (green), and squeezing estimation (gray). In-phase (in-quadrature) input squeezing in solid (dashed). For phase shift estimation (green) the curve is the same in both cases. $E_{Q}>0$ witnesses entanglement.}}
\end{figure}

In order to decide which parameter-estimation strategy is best suited to detect entanglement we show in Figure \ref{F2} \MII{the evolution of the maximal value of $E_{Q}$ when optimizing input squeezing parameters $r_A$ and $r_B$ to follow the black dashed curves in Figures \ref{F3}, \ref{F3a}, and \ref{F3b} versus amount of squeezing in dB in Alice's mode in the case of in-phase and in-quadrature input squeezing. As discussed in Sections \ref{subsec:d}, \ref{subsec:ps} and \ref{subsec:sh}, for in-phase input squeezing the maximum of $E_Q$} is obtained for $r_{A}=r_{B}$ (dashed line along the diagonal in Figures \ref{F3}a, \ref{F3a} and \ref{F3b}a). For in-quadrature input squeezing the maximum is obtained for $r_{A}=-r_{B}$ for phase-shift (dashed line along the diagonal in Figure \ref{F3a}), and for $r_{B}=\log{(1/(1+2\sinh{(r_{A})})^{1/2})}$ and $r_{B}=(-r_{A}+\log(1+e^{r_{A}}-e^{2r_{A}}))/2$ for displacement and shearing, respectively (dashed line below the diagonal in Figures \ref{F3}b and \ref{F3b}b). Remarkably, for values of input squeezing lower than $\approx 5$ dB, the best strategy is to jointly estimate the displacement (solid, blue). For larger values of input squeezing, phase shift and shearing estimation offer a greater sensitivity to entanglement (green and orange, respectively). On the contrary, as we saw above the joint estimation of the squeezing parameter does not offer any information on the entanglement of this state (solid, gray). 

In terms of resources, displacement estimation is also advantageous. Both probability distributions and quadrature variances corresponding to Alice and Bob can be directly measured with homodyne detection. Likewise, shearing estimation can be performed with homodyne detection, but fourth-order moments of the distributions (kurtosis) are necessary, which implies in general larger data sets. In the case of phase estimation, photon-number variances are necessary, which implies adding complexity to the detection.

Another great advantage of displacement estimation is that the displacement operation can be applied in post-processing: once the probability distribution $\mathcal{P}(x_A,x_B|0)$ is measured, the displaced probability distribution [under the Hamiltonian $\hat{H}_{\pm} = (\hat{p}_A \pm \hat{p}_B)/2$] is directly given by $\mathcal{P}(x_A,x_B|\theta) = \mathcal{P}(x_A+ \theta,x_B \pm \theta|0)$ \cite{Comment1}, from which one can compute the classical FI (see Sec. \ref{sec:discretization_data}) -- which we know saturates the QFI in this case, leading \DBR{to} the best possible estimation. On the contrary, the shearing and phase-shift operations can not be implemented in post-processing using just the probability distribution as full information about the state is needed for such operations. Thus, shearing and phase-shift unitaries have to be implemented at the level of the experimental setup or by post-processing after measuring the full quantum state by \DBR{Wigner tomography \cite{Ourjoumtsev2009} or Husimi Q function sampling \cite{Chabaud2021}}. In addition to this complication, contrary to the displacement operation, as we pointed out in Sec. \ref{subsec:ps} and Sec. \ref{subsec:sh}, the FI can be optimized with local rotations of the measured quadratures, but only saturates the QFI when one mixes modes $A$ and $B$.



\section{Realistic detection of non-Gaussian entanglement}\label{sec:V}

In this section we study the measurement of entanglement in a realistic scenario. As we found above, estimating displacement is the best strategy for ideal detection at moderate values of squeezing. Moreover, it is the simplest one, as the variances of the generators --field quadratures-- are directly measured with homodyne detection. We thus focus on this option in the following. A similar analysis could be carried out for shearing and phase-shift estimation. Below we analyze the effect of unbalancing the sensitivity in the displacement estimation, the effect of losses on the detection of entanglement and the discretization of the sampled data to build a joint probability distribution and calculate the Fisher information.

\subsection{Optimization of displacement axis for entanglement witness} \label{subsec:opt}




In the previous section, we analyzed the detection of entanglement through displacement estimation when displacing the input state along the axes $x_{A}=\pm x_{B}$. However, we can optimize the entanglement witness displacing the input state along an axis different to $x_{A}=\pm x_{B}$ or, in other words, unbalancing the sensitivity related to Alice and Bob in the joint parameter estimation. The idea is the following: instead of displacing the same amount $(1,\pm 1)$ in both amplitude quadratures, we displace $(\sqrt{2} \cos(\delta+\pi/4), \pm \sqrt{2} \sin(\delta+\pi/4))$ along $x_A$ and $x_B$, respectively, where $\delta \in [0, \pi]$ is an angle that we can optimize for each pair of values of $r_{A}$ and $r_{B}$. This leads to a new Hamiltonian $\hat{H}_{\pm}=(\cos(\delta+\pi/4) \hat{p}_{A}  \pm \sin(\delta+\pi/4) \hat{p}_{B})/\sqrt{2}$. Calculating the optimal entanglement witness $E_{Q}$ of Equation (\ref{Ent}) we find now
\begin{equation}\label{Ed2}
E_{Q}^{\delta}=E_{Q}\cos(2\delta).
\end{equation}
Therefore, displacing along $x_{A}=\pm x_{B}$ ($\delta=0, \pi$) is indeed the optimal strategy and displacement along any other axis can only degrade the detection of entanglement since $\vert \cos(2\delta) \vert \leq 1$.

\subsection{Optical losses}

The effect of optical losses can be entirely absorbed by the covariance matrix when it is the same in both modes \cite{Walschaers2019}. The covariance matrix of the input squeezed state $V_{0}$ is modified in the following way $V_{\eta}=(1-\eta) V_{0}+\eta\mathds{1}$, where $\eta$ represents the amount of losses. For instance, the probability distribution related to the quantum state given by Equation (\ref{UnSymPsi}) with $\phi=\pi/4$ and $r_{A}=r_{B}\equiv r$ is now 
\begin{equation} \nonumber
\mathcal{P}_{\eta}(x_{A},x_{B}) \propto  e^{ -\frac{x_{A}^{2} + x_{B}^{2}}{2 \sigma^2} } (2 \eta e^{2r}\sigma^{2}+ (1-\eta)(x_{A} + x_{B})^{2}),
\end{equation}
with $\sigma^{2}=(1-\eta) \,e^{-2r}+ \eta$. A similar but less straightforward result is obtained for general values of $\phi$, $r_{A}$ and $r_{B}$.

\begin{figure}[h!]
  \centering
    \hspace{-0.8cm} \subfigure{\includegraphics[width=0.47\textwidth]{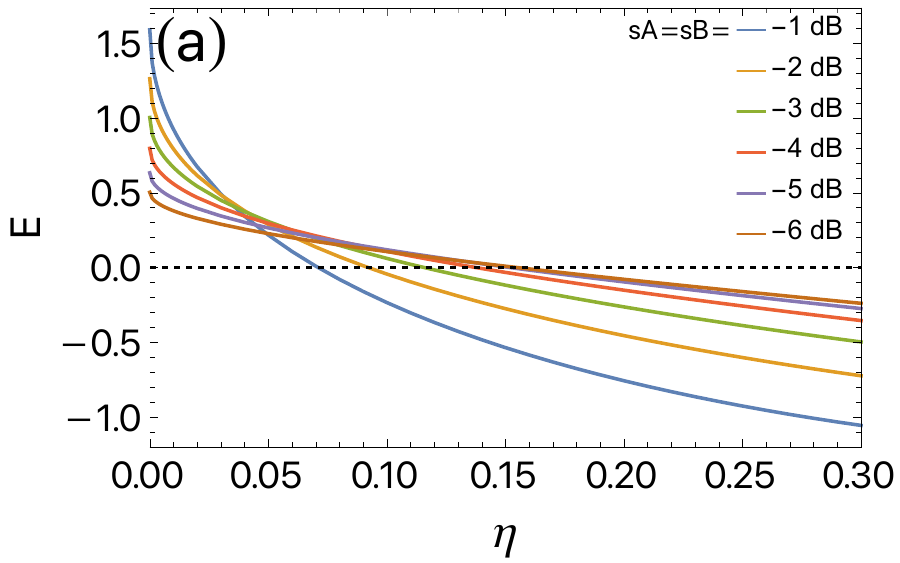}}
   \subfigure{\includegraphics[width=0.45\textwidth]{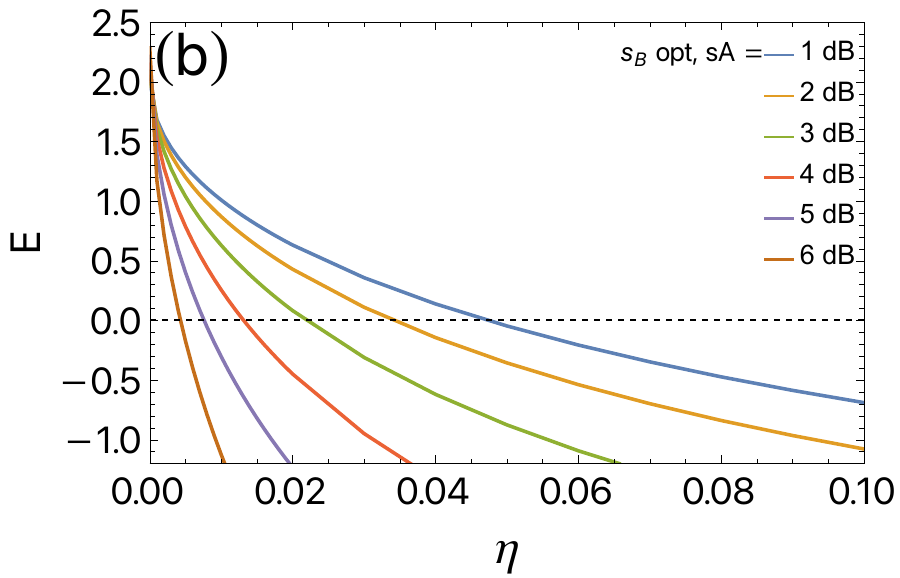}}
\vspace {0cm}\,
\hspace{0cm}\caption{\label{F4}\small{Effect of losses $\eta$ on displacement-estimation-based entanglement witness for a single photon-subtracted quantum state with $\phi=\pi/4$ and (a) optimal in-phase input squeezing $s_{A}=s_{B}<0$ (see legend), (b) optimal in-quadrature input squeezing  $s_{B}(s_{A})$ (see legend) as given by the black dashed line in Figure \ref{F3}b. $E>0$ witnesses entanglement.}}
\end{figure}

Figure \ref{F4} shows the effect of losses on the detection of entanglement for single photon-subtracted states with $\phi=\pi/4$ and different squeezing parameters: Figure \ref{F4} (a) corresponds to in-phase squeezing $s_{A}=s_{B}<0$ ranging from $-1$ to $-6$ dB; in Figure \ref{F4} (b) $s_A$ is positive, while $s_B$ is negative and equals the optimal in-quadrature squeezing -- given by  $r_{B}=\log{(1/(1+2\sinh{(r_{A})})^{1/2})}$ (see the dashed line in Figure \ref{F3}b). 
For in-quadrature input squeezing, the metrological detection of entanglement is resilient up to $\approx 5\%$, whereas for in-phase input squeezing, the metrological detection of entanglement is resilient up to $\approx 15\%$ for input values of squeezing between -4 and -6 dB. Moreover, entanglement is more resilient to losses in comparison with quantum steering, where the losses threshold is about $7\%$ for the same states \cite{Lopetegui2022}.

\begin{figure}[h!]
 \centering
   \subfigure{\includegraphics[width=0.45\textwidth]{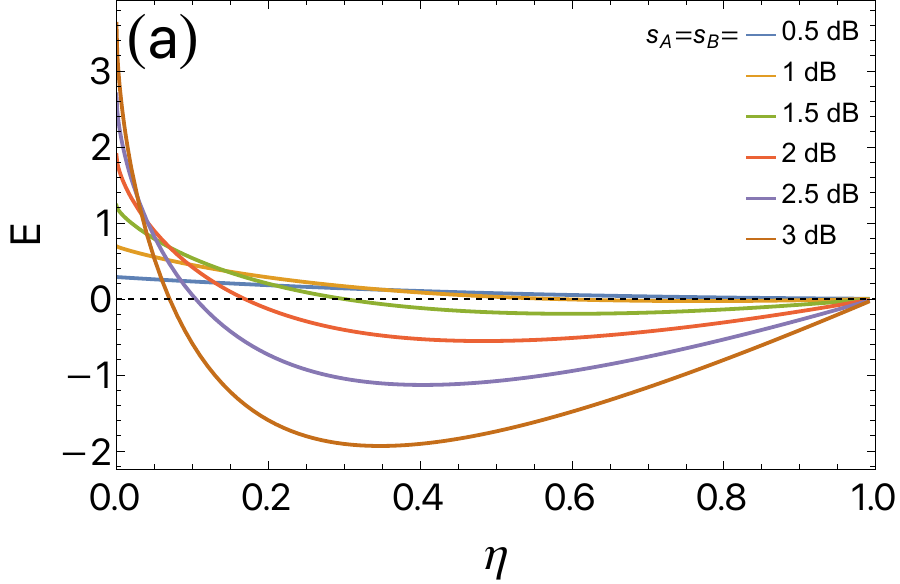}}
    \subfigure{\includegraphics[width=0.46\textwidth]{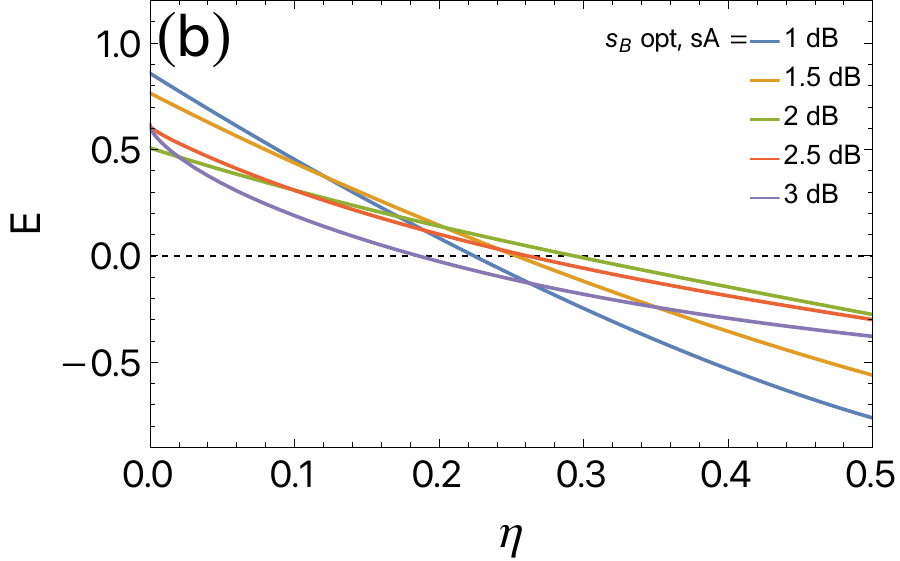}}
\caption{\label{F4b}\small{Effect of losses $\eta$ on displacement-estimation-based entanglement witness for a two photon-subtracted quantum state with $\phi=\pi/4$ and in-phase input squeezing $s_{A}=s_{B}>0$ (see legend), (b) optimal in-quadrature input squeezing $s_B(s_A)$ (see Supplemental Material). $E>0$ witnesses entanglement.}}
\end{figure}

For comparison, we show in Figure \ref{F4b} the effect of losses on the detection of entanglement for a two photon-subtracted state with $\phi=\pi/4$ \MI{in two cases: in-phase squeezing $s_{A}=s_{B}$ [Figure \ref{F4b} (a)], and optimal in-quadrature squeezing [Figure \ref{F4b} (b)]. Contrary to the one-photon subtracted case where we found the optimal value of $s_B$ for $\eta=0$ and used the same value for $\eta>0$, here we compute for each squeezing parameter $s_A>0$ and losses $\eta$ the optimal in-quadrature squeezing $s_B^{\sss \rm opt}<0$. \MII{More details on the procedure to obtain these optimal values can be found in the Supplemental Material. Table \ref{tab:my_label} gathers the values we found for the five input squeezing parameters $s_A$ of Figure \ref{F4b} (b). As we can notice from this Table, the squeezing parameter $s_B^{\sss \rm opt}$ can be very different and always larger in absolute value \DBR{than} the one found for $\eta = 0$ (see third row of Table \ref{tab:my_label}); it mainly depends on how the Fisher information is affected by losses. Large anti-squeezing leads to a better resilience of the Fisher information against losses and this explains why the optimal value for $|s_B|$ shifts to larger values. An explanation of the above statement based on the shape of probability distributions from which the Fisher information is computed can be found in the Supplemental Material.

A similar optimization can also be done for one-photon subtracted states, but it leads to too large squeezing values difficult to reach experimentally \DBR{with the currently available technology}. 
\begin{table}[h!]
    \centering
    \begin{tabular}{|c|c|c|c|c|c|}
     \hline
       $s_A$ (dB)  & 1 & 1.5 & 2 & 2.5 & 3  \\
       \hline
        $s_B^{\sss \rm opt}$ (dB) & -1.4 & -2.6 & -5 & -6.1 & -7.3  \\
         \hline
         $s_B^{\eta = 0}$ (dB) & -0.9 & -1.3 & -1.7 & -2 & -2.3  \\
         \hline
    \end{tabular}
    \caption{Optimal in-quadrature input squeezing parameters $s_B^{\sss \rm opt}$ for two-photon subtracted states. The third row corresponds to the optimal value $s_B^{\eta = 0}$ when $\eta = 0$ (no losses).}
    \label{tab:my_label}
\end{table}

}
} As seen from Figure \ref{F4b}, the metrological detection of entanglement for in-phase squeezing is resilient up to $\approx 99\%$ \MW{of losses} for low input squeezing, and larger than $\approx 25\%$ for 1.5 dB of squeezing. For in-quadrature squeezing, one can expect a resilience up to 30$\%$ of losses for $s_A = 2$ dB. Thus we can conclude that the effect of losses on the metrological detection of entanglement is both parameter and state-dependent, and that two-photon subtracted states are better suited to detect entanglement using displacement estimation in a realistic experiment.


It must be emphasised that our entanglement witness detects only entanglement related to the metrological sensitivity of the state \cite{Pezze2009}. The losses produce quantum decoherence and impair the metrological power of the quantum state. We have checked that other entanglement witness, such as the logarithmic negativity, detect entanglement in regions where our metrological witness cannot.

\subsection{Discretization of sampled data} \label{sec:discretization_data}

The FI and the variances associated to the local displacement generators $\text{Var}(\hat{p}_{A/B})$ have to be measured experimentally in order to compute the entanglement witness $E$ of Equation \eqref{Ent1}. The variances are directly obtained measuring the phase quadratures with homodyne detection. Estimating the FI experimentally from discrete outcomes, in contrast with the theoretical computation that assumes a continuum of outcomes, relies on the computation of a statistical distance --the Hellinger distance-- between a reference probability distribution and the parameter-dependent one \cite{Strobel2014}. The squared Hellinger distance between a parameter-dependent probability distribution $\mathcal{P}(x_{A},x_{B}\vert \theta)$ and a reference $\mathcal{P}(x_{A},x_{B}\vert 0)$ is defined as
\begin{align}\nonumber
&d_{H,\mathcal{P}}^{2}(\theta)\\ \nonumber
&=\frac{1}{2}\iint_{R} \left(\sqrt{\mathcal{P}(x_{A},x_{B}\vert \theta)}-\sqrt{\mathcal{P}(x_{A},x_{B}\vert 0)} \right)^2 dx_{A} dx_{B}.
\end{align}
The Taylor expansion of the squared Hellinger distance to second order yields \cite{Strobel2014}
\begin{equation}\nonumber
d_{H,\mathcal{P}}^{2}(\theta)=\frac{F}{8}\theta^{2}+\mathcal{O}(\theta^{3}),
\end{equation}
with $F\equiv F(\hat{\rho}_{A B}, \hat{H})$ the FI. Thus, a quadratic fit is enough to calculate the FI. 

However, in an experimental implementation we do not have exact probability distributions $\mathcal{P}(x_{A},x_{B}\vert \theta)$, but relative frequency distributions $\{\mathcal{F}(x_{A},x_{B}\vert \theta)\}$, which approach the probability distributions for infinitely many independent measurements. In this case, due to statistical fluctuations $\delta\mathcal{F}$, the squared Hellinger distance varies when repeating the measurement. Taking the outcome frequencies from a sample of $M$ experimental realizations, the sample average of the squared Hellinger distance \DBR{$d_{H,\mathcal{F}}^{2}(\theta)$ between two relative frequencies $\{\mathcal{F}(x_{A},x_{B}\vert \theta)\}$ and $\{\mathcal{F}(x_{A},x_{B}\vert 0)\}$ is approximately \cite{Strobel2014}}
\begin{equation}\label{Hell}
\langle d_{H,\mathcal{F}}^{2}(\theta) \rangle = c_{0} + (\frac{F}{8}+c_{2}) \theta^{2} + \mathcal{O}(\theta^{3},\delta\mathcal{F}^{3}),
\end{equation}
with $c_{0}=(n-1)/4 M$, $c_{2} \approx F(1+n)/32M$ and $n$ the number of \DBR{discrete bin pairs $\{x_{A},x_{B}\}$ with measured values.} \MII{In our case the values of $c_0$ and $c_2$ are obtained through the fitting procedure of simulated data with a quadratic function.} Note that $\langle d_{H,\mathcal{F}}^{2}(\theta) \rangle$ converges asymptotically to $d_{H,\mathcal{P}}^{2}(\theta)$ as $M\rightarrow \infty$ and hence the estimation of $F$ is asymptotically unbiased with the bias decreasing as $M^{-1}$.

\begin{figure}[t]
  \centering
    \subfigure{\includegraphics[width=0.49\textwidth]{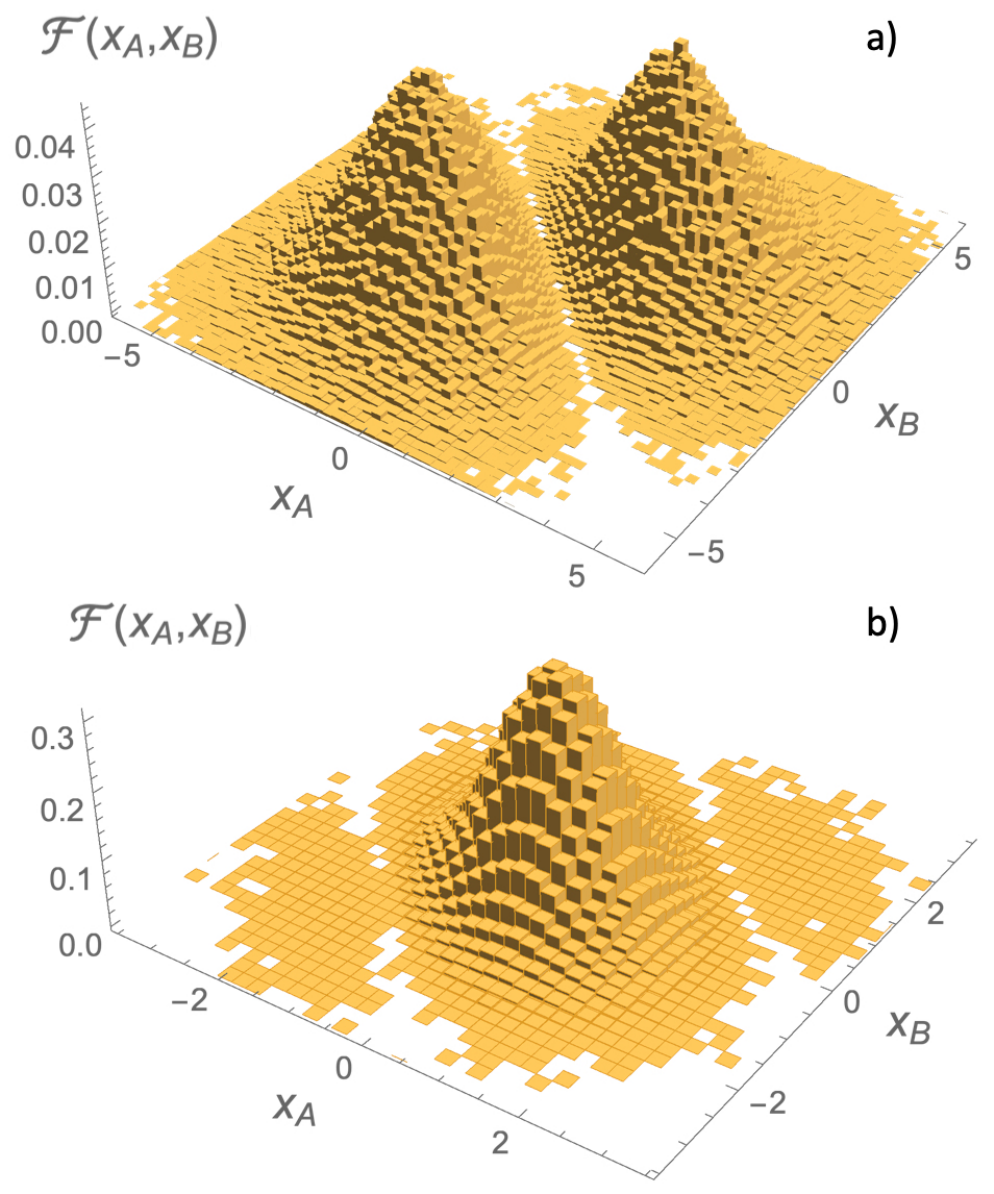}}
\hspace{0cm}\caption{\label{F5}\small{Sampled joint relative frequency distributions $\mathcal{F}(x_{A},x_{B})$ obtained through rejection sampling of photon-subtracted states. a) single-photon subtraction with $r_A=r_B=-0.5$ ($\vert s_{A/B} \vert=-4.3$ dB) and b) two-photon subtraction with $r_A=r_B=0.2$ ($\vert s_{A/B} \vert=1.73$ dB) for $5\times10^5$ samples. Bin size $\Delta$=0.2 (in units of $x_{A/B}$). }}
\end{figure}

In the following we study the protocol by simulating homodyne detection with rejection sampling of the theoretical probability distributions obtained from Equation \eqref{UnSymPsi}. We partition the real line corresponding to the outcomes of the quadrature measured by Alice and Bob in a series of bins with a given bin size $\Delta$. We consider an even number of bins as the mean value of the field is zero for our non-Gaussian probe state. Figure \ref{F5} shows two examples of sampled joint relative frequency distributions $\{\mathcal{F}(x_{A},x_{B})\}$ obtained through rejection sampling of a single photon-subtracted state probability distribution \DBR{$\mathcal{P}(x_{A},x_{B})$} given by Equation \eqref{UnSymPsi} for $r_A=r_B=-0.5$ and of a two photon-subtracted state probability distribution with $r_A=r_B=0.2$ (see Supplementary Material). The number of samples is $M=5\times10^5$ and the bin size $\Delta$=0.2 (in the units of $x_{A/B}$). 

We list below the steps to follow in order to calculate the FI: 
\begin{enumerate}
   
\item we take the two sets of sampled data corresponding to Alice $\vec{x}_{A}$ and to Bob $\vec{x}_{B}$ and split the sampled data ($\vec{x}_{A}$,$\vec{x}_{B}$) of total size $M$ in two equal sets. 

\item we bin the data in areas of given size and compute the relative frequencies \DBR{$\{\mathcal{F}(x_{A},x_{B}\vert 0)\}\approx \mathcal{P}(x_{A},x_{B}\vert 0)$} of the first set that is used as a reference. 

\item we displace the data of the second set by an amount $\theta$ --the displacement parameter--, bin the data and compute the relative frequencies \DBR{$\{\mathcal{F}(x_{A},x_{B}\vert \theta)\}\approx \mathcal{P}(x_{A},x_{B}\vert \theta)$} that are used as a probe. 

\item we calculate the square root of each relative frequency for the reference and the displaced data, take the difference and square it. 

\item we calculate the sample average of the squared Hellinger distance $\langle d_{H,\mathcal{F}}^{2}(\theta) \rangle$ for a value of $\theta$. 

\item we repeat this process for different values of $\theta$ and fit the results to a parabola, obtaining the FI with its statistical error through Equation $\eqref{Hell}$. 

\end{enumerate}

Using this value of FI and the sum of the variances of the phase quadratures we calculate the entanglement witness E through Equation \eqref{Ent1}.

\begin{figure}[h!]
  \centering
    \subfigure{\includegraphics[width=0.47\textwidth]{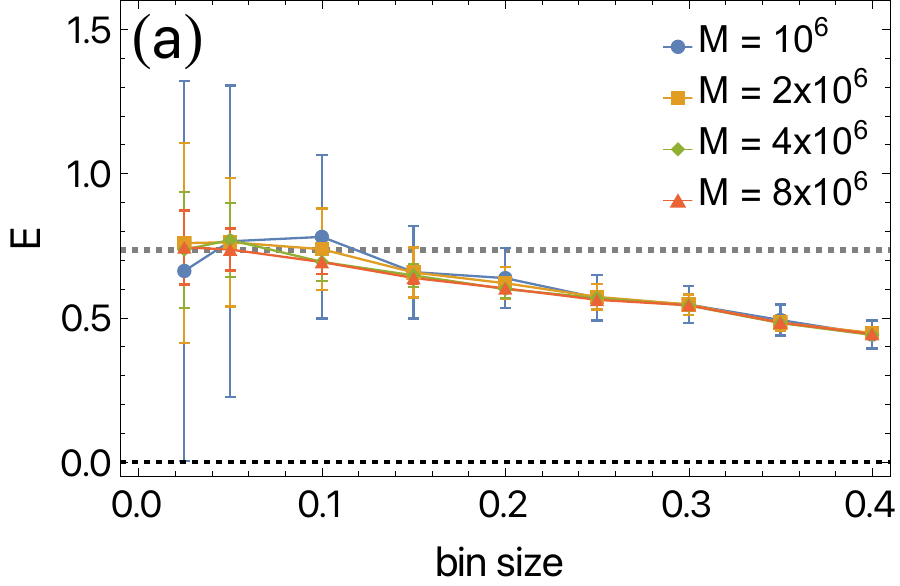}}
    \subfigure{\includegraphics[width=0.48\textwidth]{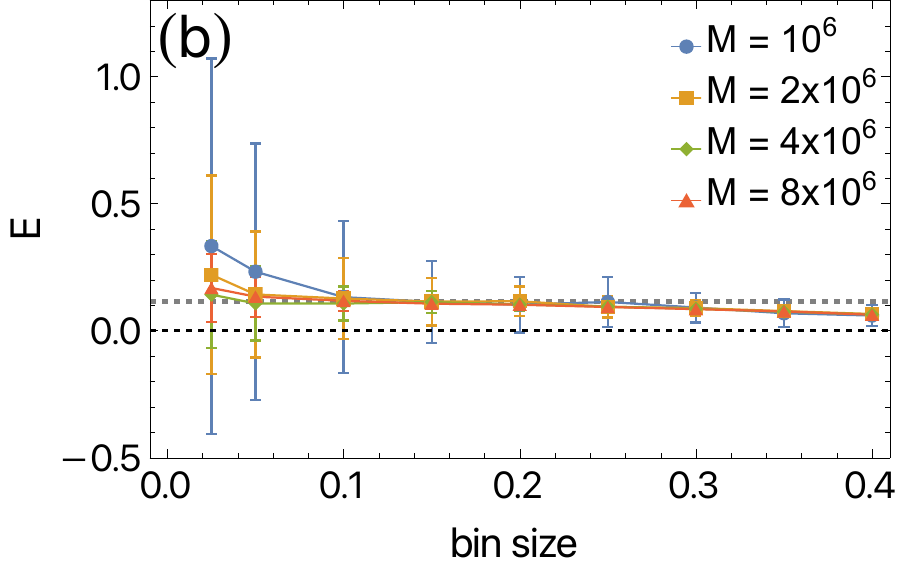}}
\vspace {0cm}\,
\hspace{0cm}\caption{\label{fig:figure11}\small{Effect of bin size $\Delta$, number of total samples $M$ and losses $\eta$ on the entanglement witness $E$ for a \MW{single-}photon-subtracted quantum state with $\phi=\pi/4$ and $r_{A}=r_{B}=-0.5$ ($\vert s_{A/B} \vert=-4.3$ dB), and (a) $\eta = 0$ or (b) $\eta = 0.1$. Averages and errors are calculated over 100 simulations. $E>0$ witnesses entanglement. The non-zero horizontal gray dashed lines correspond to the theoretical value. 
}}
\end{figure}

\begin{figure}[h!]
  \centering
    \subfigure{\includegraphics[width=0.47\textwidth]{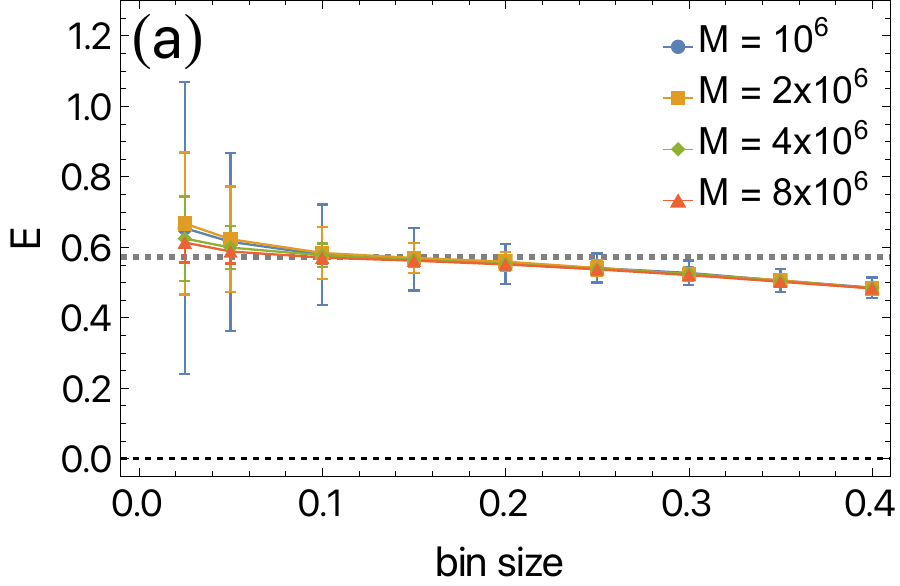}}
    \subfigure{\includegraphics[width=0.48\textwidth]{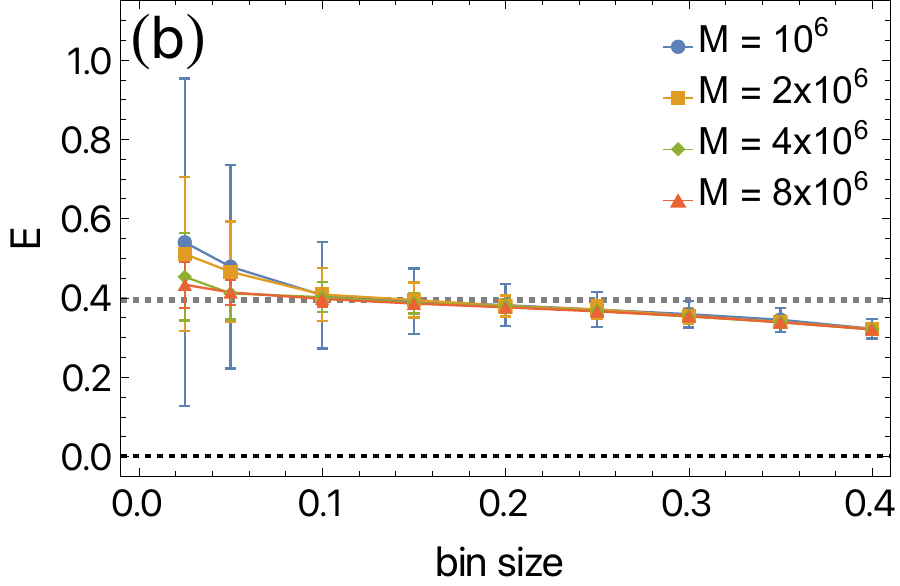}}
  \subfigure{\includegraphics[width=0.5\textwidth]{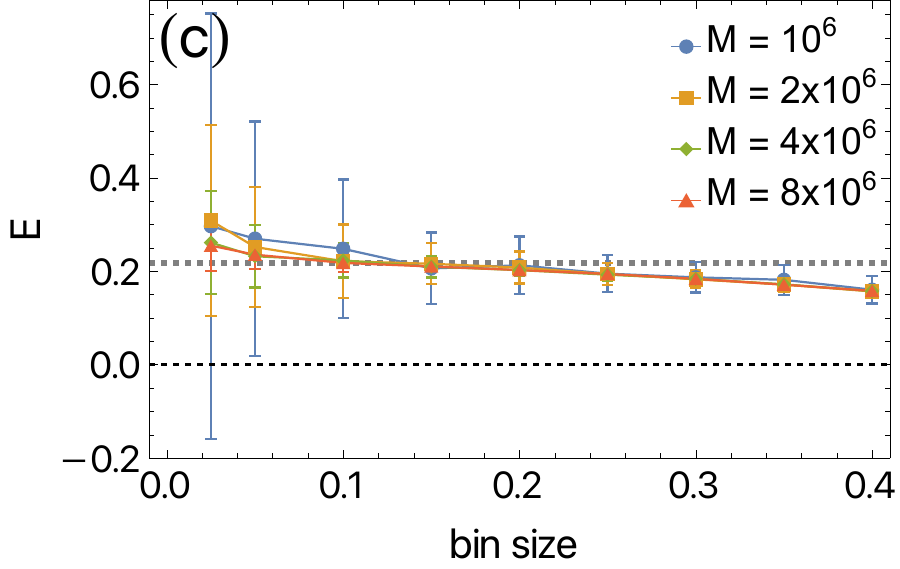}}
\caption{\label{fig:figure12}\small{Effect of bin size $\Delta$, number of total samples $M$ and losses $\eta$ on the entanglement witness $E$ for a two-photon-subtracted quantum state with $\phi = \pi/4$, $r_{A}=r_{B}=0.1$ ($\vert s_{A/B} \vert=0.87$ dB) and (a) $\eta = 0$, (b) $\eta = 0.1$, (c) $\eta = 0.25$. Averages and errors are calculated over 100 simulations. $E>0$ witnesses entanglement. The non-zero horizontal gray dashed lines correspond to the theoretical value. 
}}
\end{figure}

We show in Figures \ref{fig:figure11} and \ref{fig:figure12} the effect of data discretization and number of samples in the detection of entanglement for lossless and lossy cases.
Figure \ref{fig:figure11} corresponds to the case of a one-photon subtracted state given by Equation \eqref{UnSymPsi} with $r_A=r_B=-0.5$, either in the lossless case, i.e., $\eta = 0$ [see Figure \ref{fig:figure11} (a)], or in the lossy case [$\eta = 0.1$, see Figure \ref{fig:figure11} (b)]. 
Results for a two-photon subtracted state with $r_A=r_B=0.1$ are depicted in Figure \ref{fig:figure12} (a) (lossless case), \ref{fig:figure12} (b) (lossy case with $\eta$ = 0.1), and  \ref{fig:figure12} (c) (lossy case with $\eta$ = 0.25).
We displace our second data set of size $M/2$ between $\theta \in \{-0.05,0.05\}$ in steps of $5\times 10^{-3}$, resulting in 20 data points that we fit with a parabola using Equation \eqref{Hell}. We partition the outcome quadratures measured by Alice and Bob in a series of bins of size $\Delta$. We perform 100 simulations for each value of bin size and total number of samples to obtain statistical averages and errors. We show the value of entanglement witness $E$ obtained using a continuous probability distribution in solid gray, and the values and errors obtained for different bin size $\Delta$ and total samples $M$ in color. The colors represent different number of samples: $M=10^{6}$ (blue), $M=2\times10^6$ (orange), $M=4\times 10^6$ (green) and $M=10^{7}$ (red). 

We find that the distance between the computed value from the simulated data and the theoretical value decreases as the bin size shrinks. For large bin size, the number of samples does not affect significantly the accuracy of the measurement. However, for smaller bin size, the accuracy of the discretized estimation raises as the number of samples increases. In general, the statistical error obtained from the fit is lower as the bin size increases. Note that a discretization with an insufficient number of points can lead to an overestimation of the entanglement witness $E$. 
We find that for both one and two-photon subtracted states in the lossless case the estimation is in good agreement with the theoretical value for $M\geq 2\times10^6$ and $\Delta \leq 0.1$. In the lossy case, more samples are necessary for the same value of bin size $\Delta$ and overestimation is more significant. To not overestimate the entanglement we should use  $M\geq 2\times10^6$ and $\Delta > 0.1$. Notably, in both cases we detect entanglement even using a coarse-grained bin size $\Delta=0.4$ and a relatively low number of samples $M=10^6$.

\begin{figure}[h!]
  \centering
  \includegraphics[width=0.45\textwidth]{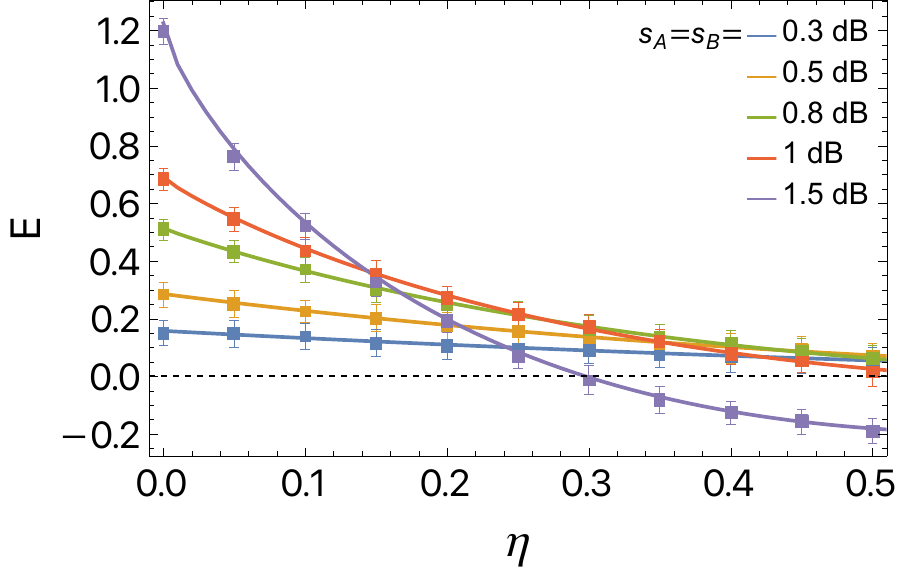}
\caption{\label{fig:figure13}\small{Displacement-estimation-based entanglement witness $E$ for a two-photon subtracted state with $\phi = \pi/4$, $s_A = s_B$, $s_A$ ranging from 0.3 dB to 1.5 dB (see legend). The square with error bars correspond to simulated sampled data calculated over 100 simulations with bin size $\Delta = 0.15$ and total samples $M=2$ millions. The solid curves correspond to the expected theoretical results.}}
\end{figure}

In Figure \ref{fig:figure13} we show that the above method is working well for a wide range of two-photon subtracted states with various squeezing parameters. The square with error bars are directly computed from the sampled data with $\Delta = 0.15$ and $M = 2\times10^6$ and all agree with the theoretical entanglement witness $E$. It makes it possible to detect non-Gaussian entanglement up to 45$\%$ of losses for squeezing parameters up to 1 dB, and up to 30$\%$ of losses if we include the 1.5 dB case.  

\begin{figure}[h!]
  \centering
  \includegraphics[width=0.45\textwidth]{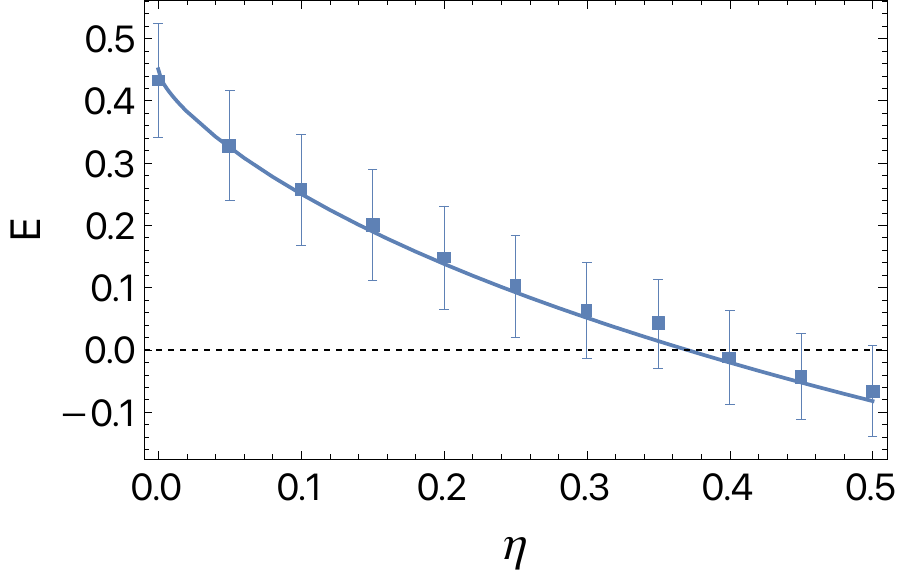}
\caption{\label{fig:figure13bis}\small{Squeezing-estimation-based entanglement witness $E$ for a two-photon subtracted state with $\phi_A = -\phi_B=\pi/4$, $r_A = -0.1$ ($s_A = -0.87 $ dB), $r_B=-0.2$ ($s_B = -1.74$ dB). The square with error bars correspond to simulated sampled data calculated over 100 simulations with bin size $\Delta = 0.2$ and total samples $M=2$ millions. The solid curves correspond to the expected theoretical results.}}
\end{figure}

\MI{In the above paragraphs we studied cases where the displacement-based metrological criterion works well, but for which usual Gaussian witnesses are also able to detect entanglement for two-photon subtracted states and for certain values of input squeezing parameters. Therefore, in the following, we would like to emphasize a case where Gaussian witnesses do not work, but the metrological criterion does. As discussed in Section \ref{sec:III}, when we subtract two photons in orthogonal modes, quantum correlations cease to appear at the level of the covariance matrix, making second-order-moment-based entanglement criteria inefficient. Likewise, it turns out that for these type of states the first-order metrological criterion (i.e., using displacement generators) will also fail; thus, it means that we need to consider second order generators. In this case, we can use squeezing operators for which $E_Q \neq 0$ for two-photon subtracted states. The interest of using these generators is two-fold here: (1) the squeezing operation onto the state amounts to squeeze the position quadratures of modes $A$ and $B$ by a factor $e^\theta$ or $e^{-\theta}$ (see Section \ref{sec:subsecIVD}). Therefore, once the joint probability distribution $P(x_A, x_B|0)$ is measured experimentally, the transformation can be easily implemented in post-processing to obtain $P(x_A, x_B|\theta)$; (2) the FI saturates the QFI when considering the joint probability distributions in the basis $(x_A, x_B)$. 

For all the reasons explained above, we applied the squeezing-estimation-based entanglement on a two photon subtracted states with $r_A = -0.1 $ ($s_A = -0.87 $ dB) and $r_B =-0.2$ ($s_B = -1.74$ dB), $\phi_A = \pi/4$ and  $\phi_B = -\pi/4$, where $\phi_A$ and $\phi_B$ control the modes where we subtract the photons. As shown in Fig. \ref{fig:figure13bis}, with a bin size $\Delta = 0.2$ and a number of samples $M = 2 \times 10^{6}$, there is a good agreement between the theoretical entanglement value $E$ and the entanglement estimation based on simulated sampled data calculated over 100 simulations. To obtain these results we also needed an estimation of the local variances of the squeezing generators that can be directly obtained from the statistics of the sampling data using the kurtosis in different quadrature bases (see Section IV of the Supplemental material for more details). One sees that in this case we can detect entanglement up to almost 30$\%$ of losses.

 }

\section{Experimental implementations} \label{sec:VI}

Let us discuss possible practical implementations of this protocol. There are few approaches depending on the degree of freedom --or mode-- selected to encode the quantum information: path, polarization, frequency and so on. The shared feature of the input modes is that they are independent and excited in squeezed states. An event measured by a single-photon detector fed by a small fraction of power from Alice and Bob's modes where which-mode information is erased, heralds the subtraction of a photon delocalized between the two modes \cite{Ourjoumtsev2009}. Two balanced homodyne detectors with a common local oscillator LO retrieve then the joint probability distribution. The sketch of Figure \ref{F0} is pretty accurate for path-encoded modes where a common beam splitter erases the which-path information. 

In the case of spectral modes where the number of modes is usually larger than two --for instance in a multimode frequency-comb Gaussian resource \cite{Roslund2013, Cai2017}-- mode-selective photon-subtraction is accomplished by sum-frequency generation \cite{Ra2020}. The detection of an up-converted photon heralds the subtraction of a photon from a multimode input state in one or various spectral modes selected by a pump suitably tailored in frequency. The joint probability distribution of photon-subtracted spectral modes can be retrieved by spectrally-resolved homodyne detection \cite{Ansquer2021}. This approach allows to measure simultaneously the quadratures of the electric field in a number of frequency-band modes. Then, applying a change of basis between the photon-subtracted spectral modes and these frequency-band modes one retrieves the quadrature traces in the modes of interest and hence, the joint probability distribution.

Moreover, we outline that in an experiment, in order to prove that the entanglement results entirely from the non-local photon subtraction, one would use the data from the unconditioned state to test our entanglement witness and demonstrate the independence of the two input squeezed states. 

Finally, comparing our simulations with the values measured by Y.-S. Ra et al. \cite{Ra2020} for one-photon subtracted states, where the squeezing of the first and second spectral modes is $s_{A}=-2.3$ dB and $s_{B}=-1.7$ dB, respectively, with purities above $90\%$ and detection losses of the order of $12\%$, and with the values obtained in T. Takanashi et al. \cite{Takanashi2008} for two-photon subtracted states, with $-3.0$ dB of squeezing and detection losses of $15\%$, we conclude that with a reasonable number of samples ($\approx 10^6$) it is possible to witness non-Gaussian entanglement using exclusively homodyne detection with an experimentally feasible protocol.

\section{Conclusions and outlook}\label{sec:VII}

We proposed a protocol based on Fisher information for witnessing entanglement in an important class of non-Gaussian states: photon-subtracted CV states. The protocol is based on the metrological entanglement criterion proposed in \cite{Gessner2016}, and its strength comes from its simplicity, as it relies solely on homodyne detection. Our approach witnesses entanglement not detected by Gaussian criteria, like for instance Duan {\it et al.} criterion, using the same resources, i.e. quadrature measurements.

We characterized the optimal metrologically-useful entanglement of single- and two-photon-subtracted states analyzing their metrological power in estimation of parameters generated by all single-mode Gaussian gates, namely: displacement, phase shift, shearing and squeezing. We analyzed displacement estimation in details since it gives the largest sensitivity for currently experimentally-relevant values of squeezing ($\leq 5$ dB) and it can be applied in postprocessing, thus minimizing the resources necessary in non-Gaussian entanglement characterization and outperforming other protocols where quantum-state tomography is needed. 

We demonstrated that our protocol is relevant and experimentally feasible using data from a simulated experiment where the effect of losses, data discretization, and number of samples were taken into account. Our results show that non-Gaussian entanglement can be detected with a feasible number of measurements and data binning. It is well known that losses impair the metrological power of quantum states. However, we found that our metrology-based entanglement detection is resilient up to $30\%$ losses for some purely non-Gaussian entangled states. 

The general setup of Figure~\ref{F0} is versatile and can be used to both implement Gaussian entanglement detection protocols based on the covariance matrix and our metrological protocol for non-Gaussian entanglement. For certain classes of states, we believe that this should be sufficient to be able to detect entanglement in any mode basis. However, to determine whether or not a state is passively separable, as would be required for the sampling protocols in \cite{Chabaud2022}, one would still need to certify the presence of entanglement in every possible mode basis. While our work certainly offers us a useful experimental tool, we also hope that it will be a step towards finding new techniques that allow us to certify entanglement in every possible mode basis. After all, non-Gaussian entangled states encompass a huge state space and we have just started to \DBR{scratch its surface}. In order to gain insight about general features of this exotic quantum feature, in future work we will analyze entangled states based on other non-Gaussian operations and connect our entanglement criterion with others based on higher-order covariance matrices \cite{Zhang2021, Zhang2023}.

\section*{Acknowledgements}
This work received funding from the ANR JCJC project NoRdiC (ANR-21-CE47-0005), the European Union’s Horizon 2020 research and innovation programme under Grant Agreement No. 899587, and the QuantERA II project SPARQL that has received funding from the European Union’s Horizon 2020 research and innovation programme under Grant Agreement No 101017733. This work was also funded by MCIN/AEI/10.13039/501100011033 and the European Union “NextGenerationEU” PRTR fund [RYC2021-031094-I], by the Ministry of Economic Affairs and Digital Transformation of the Spanish Government through the QUANTUM ENIA project call - QUANTUM SPAIN project, by the European Union through the Recovery, Transformation and Resilience Plan - NextGenerationEU within the framework of the Digital Spain 2026 Agenda, and by the CSIC Interdisciplinary Thematic Platform (PTI+) on Quantum Technologies (PTI-QTEP+). \DBR{This work was supported in part by the Valencian Government grant with reference number CIAICO/2021/184.} It was carried out during the tenure of an ERCIM ‘Alain Bensoussan’ Fellowship Programme.

\section*{}

\end{document}


\preprint{APS/123-QED}

\title{Supplementary material for Metrological detection of entanglement generated by non-Gaussian operations} 

\author{David Barral} 
\email{david.barral@lkb.ens.fr}
\affiliation{Laboratoire Kastler Brossel, Sorbonne Université, CNRS, ENS-PSL Research University, Collège de France, 4 place Jussieu, F-75252 Paris, France}
\author{Mathieu Isoard} 
\affiliation{Laboratoire Kastler Brossel, Sorbonne Université, CNRS, ENS-PSL Research University, Collège de France, 4 place Jussieu, F-75252 Paris, France}
\author{Giacomo Sorelli} 
\affiliation{Laboratoire Kastler Brossel, Sorbonne Université, CNRS, ENS-PSL Research University, Collège de France, 4 place Jussieu, F-75252 Paris, France}
\affiliation{Fraunhofer IOSB, Ettlingen, Fraunhofer Institute of Optronics, System Technologies and Image Exploitation, Gutleuthausstr. 1, 76275 Ettlingen, Germany}
\author{Manuel Gessner} 
\affiliation{
Departamento de Física Teórica, IFIC, Universidad de Valencia-CSIC, C/ Dr. Moliner 50, Burjassot, Valencia  46100, Spain}
\author{Nicolas Treps} 
\affiliation{Laboratoire Kastler Brossel, Sorbonne Université, CNRS, ENS-PSL Research University, Collège de France, 4 place Jussieu, F-75252 Paris, France}
\author{Mattia Walschaers}
\email{mattia.walschaers@lkb.upmc.fr}
\affiliation{Laboratoire Kastler Brossel, Sorbonne Université, CNRS, ENS-PSL Research University, Collège de France, 4 place Jussieu, F-75252 Paris, France}

\date{December 5, 2023}
\maketitle

In this supplementary material, we first compute the general expression of $E_Q$ for one-photon subtracted states (see Sec. \ref{sec:I}). Then, still for one-photon subtracted states, we discuss in Sec. \ref{sec:II} the possibility to optimize the value of the Fisher information (FI) -- to try to get closer from the Quantum Fisher information (QFI) upper bound in the case of the shearing and phase shift operators (see also Secs. IV B and IV C of the main paper).

In Secs. \ref{sec:III} and \ref{sec:IV}, we turn our attention to two-photon subtracted states. We show the pattern of a typical joint probability distribution for these subtracted states and we compute the optimal entanglement $E_{Q}$ in Sec. \ref{sec:IV} for the different generators considered in the main paper (displacement, shearing, phase-shift and squeezing). Then, in Sec. \ref{sec:V}, we discuss the optimal in-quadrature squeezing $s_B<0$ (as a function of $s_A>0$) to choose to maximize the displacement-estimation entanglement criterion in the presence of losses. In Sec. \ref{sec:VI} we show that the local variance of the squeezing operator $\hat H = (\hat{x} \,\hat{p} + \hat{p} \,\hat{x})/4$ that we need to evaluate the squeezing-based entanglement estimation (see Sec. V C of the main paper) can be obtained from experimental data using the kurtosis in different bases.


\section{General expression of $E_{Q}$ for one-photon subtracted states}\label{sec:I}

\subsection{Displacement operator}

In this section, we derive the general expression of $E_{Q}$ for a given angle $\phi$ (which controls the probability of subtraction in each mode) and for two different squeezing parameters $r_A$ and $r_B$ when the Hamiltonian is given by $(\hat{p}_A \pm \hat{p}_B)/2 $. 

By performing a change of basis from ($x_A, p_A, x_B, p_B$) to a new set of coordinates ($x^{\prime}_A, p^{\prime}_A, x^{\prime}_B, p^{\prime}_B$), it is possible to map the general wavefunction 
\begin{equation} 
\Psi(x_{A},x_{B})\equiv \langle x_{A}, x_{B} \vert\Psi\rangle \propto  e^{ -\frac{e^{2 r_{A}} x_{A}^{2} + e^{2 r_{B}} x_{B}^{2}}{4} } ((e^{2 r_{A}}-1) \cos{(\phi)} x_{A} + (e^{2 r_{B}}-1) \sin{(\phi)} x_{B})
\end{equation}
to the symmetric case with equal squeezing parameters, i.e.,
\begin{equation} \label{SymPsi_app}
\Psi(x^{\prime}_A,x^{\prime}_B) \propto  e^{ -e^{2 r} \frac{ {x^{\prime}_{A}}^{2} + {x^{\prime}_{B}}^{2}}{4} } \, ( x^{\prime}_{A} + x^{\prime}_{B}),
\end{equation}
where $r \equiv (r_A + r_B)/2$. This change of basis consists of two operations \begin{equation} \label{transf_x}
     \begin{pmatrix}
    x^{\prime}_{\rm \scriptscriptstyle A} \\
    x^{\prime}_{\rm \scriptscriptstyle B}
    \end{pmatrix}
    =
   \mathcal{R}(z) \,  \mathcal{S}(s)
    \begin{pmatrix}
   x_{\rm \scriptscriptstyle A} \\
    x_{\rm \scriptscriptstyle B}
    \end{pmatrix},
\end{equation}
with 
\begin{equation} \label{trans_D}
 \mathcal{S}(s) =     
 \begin{pmatrix}
    e^s & 0 \\
   0 & e^{-s}
    \end{pmatrix}, \quad s \equiv \frac{r_A - r_B}{2},
    \quad \text{and} \quad
     \mathcal{R}(z) =     
 \begin{pmatrix}
    \cos z & \sin z \\
    - \sin z & \cos z
    \end{pmatrix},
\end{equation}
where 
\begin{equation}
    \begin{split}
        \cos z & = \frac{ \sinh r_{\rm \scriptscriptstyle A} \, \cos \phi + \sinh r_{\rm \scriptscriptstyle B} \, \sin \phi }{\sqrt2 \, \sqrt{ \sinh^2 r_{\rm \scriptscriptstyle A} \cos^2 \phi + \sinh^2 r_{\rm \scriptscriptstyle B} \sin^2 \phi  }},\\
        \sin z & =  \frac{ - \sinh r_{\rm \scriptscriptstyle A} \, \cos \phi + \sinh r_{\rm \scriptscriptstyle B} \, \sin \phi }{\sqrt2 \, \sqrt{ \sinh^2 r_{\rm \scriptscriptstyle A} \cos^2 \phi + \sinh^2 r_{\rm \scriptscriptstyle B} \sin^2 \phi  }} .
    \end{split}
\end{equation}
Then, using the integral 
\begin{equation}
    \label{mean_xp_SupMat}
\langle \hat x_{i}^{l} \hat p_{j}^{m} \rangle=
(-2i)^{m}\iint_{\mathcal{R}} x_{i}^{l} \Psi (x_{A},x_{B})^{*}\, \frac{\partial^{m}\Psi(x_{A},x_{B})}{\partial x_{j}^{m}} dx_{A} dx_{B},
\end{equation}
and the change of coordinates \eqref{transf_x}, one finds that a symplectic transformation connects the second-order moments associated to the non-symmetric and symmetric case:
\begin{equation} \label{app:symp_trans}
    \sigma = \Lambda \, \sigma^{\prime} \, \Lambda^{\rm \scriptscriptstyle T},
    \quad \text{with} \quad
    \sigma = 
    \begin{pmatrix}
        \langle \hat{p}_{\rm \scriptscriptstyle A}^2 \rangle & \langle \hat{p}_{\rm \scriptscriptstyle A} \hat{p}_{\rm \scriptscriptstyle B} \rangle \\
        \langle \hat{p}_{\rm \scriptscriptstyle A} \hat{p}_{\rm \scriptscriptstyle B} \rangle & 
         \langle \hat{p}_{\rm \scriptscriptstyle B}^2 \rangle
    \end{pmatrix}, \,  \sigma^{\prime} = 
    \begin{pmatrix}
        \langle {{\hat{p}}^{\prime}}_{\rm \scriptscriptstyle A}{}^2 \rangle & \langle {\hat{p}}^{\prime}_{\rm \scriptscriptstyle A} {\hat{p}}^{\prime}_{\rm \scriptscriptstyle B} \rangle \\
        \langle \hat{p}^{\prime}_{\rm \scriptscriptstyle A} \hat{p}^{\prime}_{\rm \scriptscriptstyle B} \rangle & 
         \langle {\hat{p}^{\prime}}_{\rm \scriptscriptstyle B}{}^2 \rangle
    \end{pmatrix},
    \quad \text{and} \quad
     \Lambda = \mathcal{S}(s)  \mathcal{R}^{-1}(z).
\end{equation}
In particular, one has
\begin{equation}
   \langle \hat{p}_{\rm \scriptscriptstyle A} \hat{p}_{\rm \scriptscriptstyle B} \rangle =  \cos (2z) \langle  \hat{p}^{\prime}_{\rm \scriptscriptstyle A}  \hat{p}^{\prime}_{\rm \scriptscriptstyle B} \rangle 
        + \sin(2z) \left( \langle \hat{p}^{\prime}_{\rm \scriptscriptstyle A}{}^2 \rangle - \langle \hat{p}^{\prime}_{\rm \scriptscriptstyle B}{}^2 \rangle\right).
\end{equation}
Given that $\langle {\hat{p}^{\prime}}_{\rm \scriptscriptstyle A}{}^2 \rangle = \langle {\hat{p}^{\prime}_{\rm \scriptscriptstyle B}}{}^2 \rangle $, we finally find the simple relation
\begin{equation} \label{E_Q_appA}
    E_Q = \cos (2z) \, E^{\prime}_Q = \pm 2 \, e^{r_{A} + r_{B}} \, \cos (2z),
    \quad \text{with} \quad
     \cos (2z) =  \frac{\sinh z_{\rm \scriptscriptstyle A} \, \sinh z_{\rm \scriptscriptstyle B} \sin (2 \phi)}{\sinh^2 z_{\rm \scriptscriptstyle A} \, \cos^2 \phi + \sinh^2 z_{\rm \scriptscriptstyle B} \,  \sin^2 \phi}.
\end{equation}
In the symmetric case $\pi/4$ and different squeezing parameters, one finds exactly expression (\ref*{mp-Ed}) of the main text with $\epsilon = 2 z$.

Note that if we displace $(\sqrt2 \, \cos (\delta + \pi/4),  \pm \sqrt2 \, \sin (\delta + \pi/4))$ along $x_{\rm \scriptscriptstyle A}$, $x_{\rm \scriptscriptstyle B}$, Eq. \eqref{E_Q_appA} becomes
\begin{equation} 
    E_Q(\delta) \equiv \pm 2 \, e^{r_{A} + r_{B}} \, \cos (2 z) \cos (2 \delta).
\end{equation}
One finds that in the general case $E_Q(\delta)$ is maximized when $\delta =  \phi + (n-1/4)\pi, \, n \in \mathbb{Z}$, where Sgn is the sign function. We recover for $\phi = \pi/4$ the result discussed in Section V A of the main paper, i.e., that the optimal displacement is $\delta = (0, \pi)$.

\subsection{Shearing and phase shift operators}

The mapping between the non-symmetric and symmetric case through transformations \eqref{trans_D} can also be used to compute the entanglement witness $E_Q$ in the case of the shearing and phase shift operators. Here, one cannot use directly the symplectic transformation \eqref{app:symp_trans} since the expression of $E_Q$ involves higher order moments. However, one can still insert the change of variables \eqref{transf_x} in Eq. \eqref{mean_xp_SupMat}.

\section{Optimizing the choice of quadratures for Alice and Bob} \label{sec:II}
 
Alice and Bob can \textit{a priori} measure the joint probability distribution in any basis $(\xi_A, \xi_B) = (\cos \phi_A x_A - \sin \phi_A p_A$, \, $\cos \phi_B x_B - \sin \phi_B p_B)$ [as defined in the main text, see Section II]. The question that is answered in this section is: what is the the optimal choice for the angles $(\phi_A, \phi_B)$ -- for each of the three operators considered in this paper -- to maximize the Fisher information, and thus the entanglement witness $E$ [see Eq. (4) of the main paper]? 

The result is actually straightforward for the displacement estimation. As stated in the main text, the FI saturates the QFI when measuring the joint probability distribution in the plane $(x_A, x_B)$. Therefore, we only treat below the more complicated cases of the shearing and the phase shift operators.

\subsection{Shearing operator}

Fig.~\ref{fig:Sh-PhSM}(a) shows the FI as a function of both angles $\phi_A$ and $\phi_B$ for $r = r_A = r_B = -0.2$ and $\phi = \pi/4$. The red dot pinpoints the maximal value ($\simeq$ 5.89) obtained in this case for $\phi_A = 7 \, \pi/20 $ and $\phi_B = 13 \, \pi/20 $. The expected QFI for this squeezing parameter should be $3 \exp (4 \, r) \simeq$ 6.67. Therefore, contrary to the displacement estimation, it is not possible to saturate the QFI with local rotations of the quadratures. To generalize this result, we found numerically the maximal value reached by the FI for a large range of squeezing parameters [from $s = s_A = s_B = 0$ dB to $s\simeq 6$ dB -- see Fig.~\ref{fig:FI-QFI-SM}(a)]. It is clear that, here again, the FI does not saturate the QFI (dashed blue curve) whatever the squeezing parameter is.

\subsection{Phase shift operator}

\begin{figure}[h]
  \centering
    \subfigure{\includegraphics[width=0.45\textwidth]{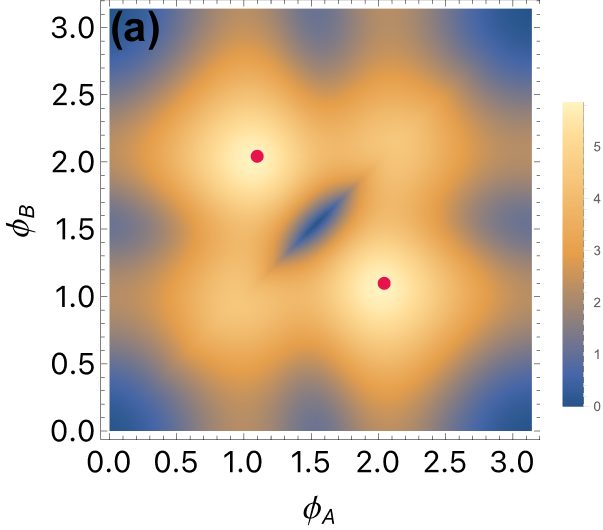}}
    \subfigure{\includegraphics[width=0.45\textwidth]{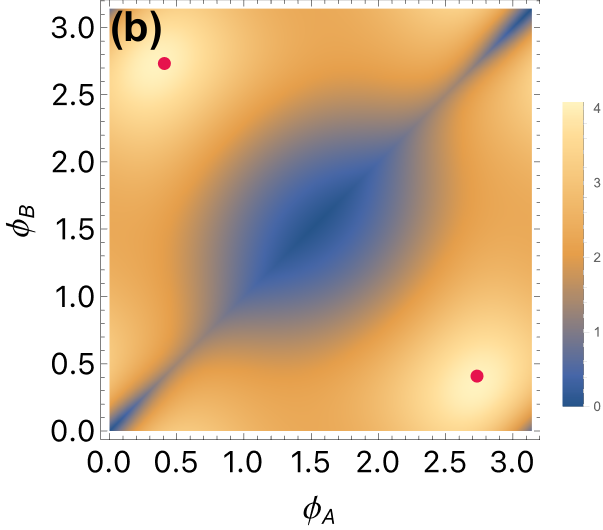}}
\vspace {0cm}\,
\hspace{0cm}\caption{\label{fig:Sh-PhSM} FI computed in the case of (a) the shearing operator for $r_A = r_B=-0.2$, $\phi = \pi/4$, (b) the phase-shift operator for $r_A = r_B=0.2$, $\phi = \pi/4$, and for different angles $\phi_A$ and $\phi_B$. These angles control the local rotations of the quadratures (see text). The red dots indicate the maximal achievable value for the FI.}
\end{figure}

\begin{figure}[h]
  \centering
    \subfigure{\includegraphics[width=0.45\textwidth]{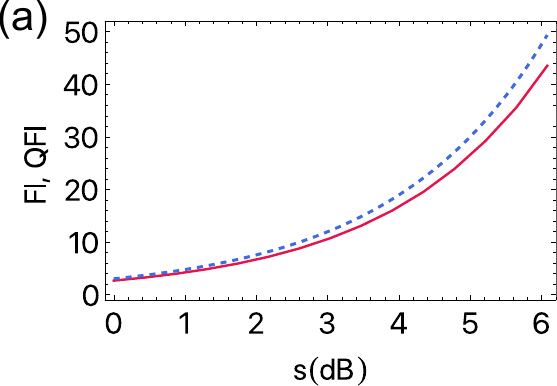}}
    \subfigure{\includegraphics[width=0.45\textwidth]{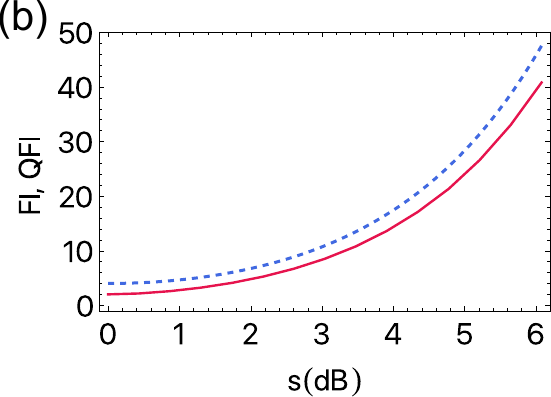}}
\vspace {0cm}\,
\hspace{0cm}\caption{\label{fig:FI-QFI-SM} Maximal value for the FI (red) in the case of (a) the shearing operator for different squeezing parameters $r = r_A = r_B<0$, and (b) the phase shift operator for $r = r_A = r_B>0$. The dashed blue curve represents the  QFI in the case of both operators. }
\end{figure}

We reproduce here the same procedure as in the previous section for the phase shift operator. Fig.~\ref{fig:Sh-PhSM}(b) shows the FI as a function of both angles $\phi_A$ and $\phi_B$ for $r = r_A = r_B = 0.2$ and $\phi = \pi/4$. The red dot pinpoints the maximal value ($\simeq$ 4.1) obtained in this case for $\phi_A = 13 \, \pi/100 $ and $\phi_B = 87 \, \pi/100 $. The expected QFI for this squeezing parameter should be $2 \cosh^2(2 r) + [-3 + 5 \cosh (4 r)] \simeq$ 6.02. Therefore, here again, it is not possible to saturate the QFI with local rotations of the quadratures. As in the previous section, Fig.~\ref{fig:FI-QFI-SM}(b) shows the maximal value reached by the FI for a large range of squeezing parameters (from $s = s_A = s_B = 0$ dB to $s\simeq 6$ dB); the FI does not saturate the QFI (dashed blue curve) whatever the squeezing parameter is.

\section{Two-photon subtracted state}\label{sec:III}

Let us consider two independent single-mode squeezed states respectively related to Alice and Bob
\begin{equation}\label{Sq}
\vert \Psi_{0}\rangle =\hat{S}_{A}(r_{A},\theta_{A}) \hat{S}_{B}(r_{B}, \theta_{B}) \vert 0 0 \rangle,
\end{equation}
 where $\hat{S}_{I}(r_{I})=\text{exp}\{(-r_{I}/2)(\hat{a}_{I}^2 -\hat{a}_{I}^{\dagger 2} )\}$ is the single-mode squeezing operator, and $r_{I} \in \mathbb{R}$ is the squeezing parameter for each mode $I=A,B$. The amount of squeezing in decibels is given by $s_{I}=10 \log_{10} (e^{-2 r_{I}})$.

Next, we perform a delocalized subtraction of two photons on this state. This operation produces a superposition of two-mode squeezed vacuum and squeezed single-photon states that one can show that up to normalizations is 
\begin{align}
&\vert\Psi\rangle \propto (\cos(\phi) \hat{a}_{A}+\sin(\phi) \hat{a}_{B})^{2} \vert \Psi_{0}\rangle = \\ \nonumber
&\hat{S}_{A}(r_{A}) \hat{S}_{B}(r_{B}) \sum_{k=0}^{2}
\begin{pmatrix}
2 \\ k 
\end{pmatrix}
(\cos(\phi) \hat{U}_{A})^{2-k}(\sin(\phi) \hat{U}_{B})^k \vert 0 0\rangle,
\end{align}
where $\hat{U}_{I=A,B}\equiv\hat{S}_{I}^{\dag}(r_{I})\hat{a}_{I}\hat{S}^{}_{I}(r_{I})=\cosh(r_{I})\hat{a}_{I}+\sinh(r_{I}) \hat{a}^{\dag}_{I}$. The parameter $\phi$ controls the probability of subtraction in each mode and we have considered in-phase subtraction. The wavefunction of a two-photon subtracted state in the amplitude quadratures of the optical field is given by
\begin{align} \label{UnSymPsi}\nonumber
&\Psi(x_{A},x_{B})\equiv \langle x_{A}, x_{B} \vert\Psi\rangle \propto  e^{ -\frac{e^{2 r_{A}} x_{A}^{2} + e^{2 r_{B}} x_{B}^{2}}{4} } \times \\
&\hspace{-0.2cm}[4(1-e^{2r_A} \cos^2(\phi)-e^{2 r_B} \sin^2(\phi)) +2(e^{2 r_{A}}-1)^{2} \cos^2(\phi) x_{A}^2+2(e^{2 r_{B}}-1)^{2}  \sin^2(\phi)x_{B}^2+ 2(e^{2 r_{A}}-1)(e^{2 r_{B}}-1)\sin(2\phi) x_{A} x_{B}].
\end{align}

\begin{figure}[h]
  \centering
    \subfigure{\includegraphics[width=0.45\textwidth]{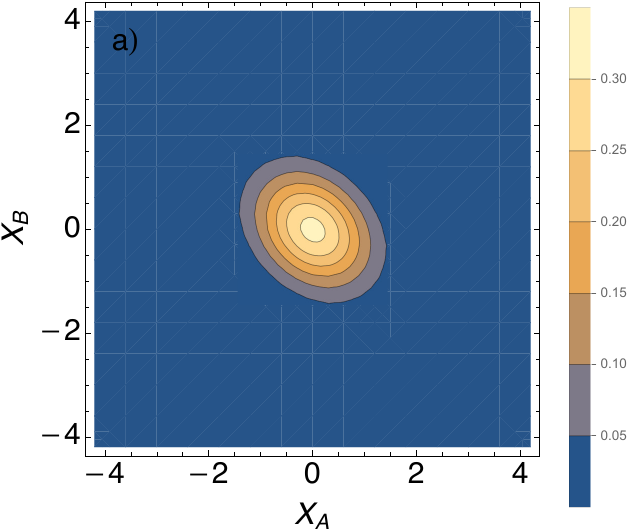}}
    \subfigure{\includegraphics[width=0.45\textwidth]{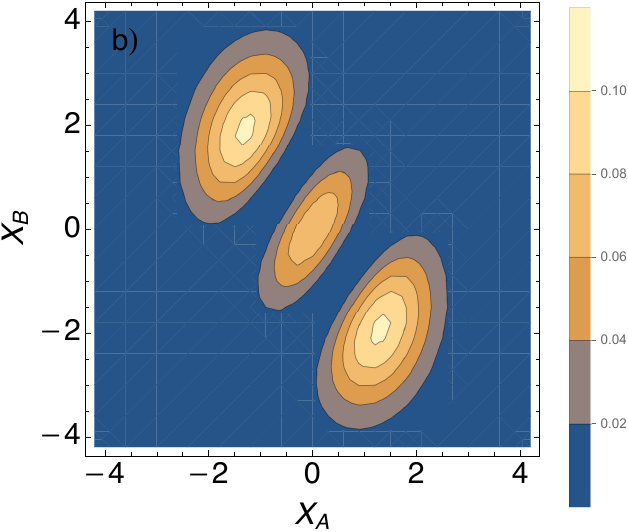}}
\vspace {0cm}\,
\hspace{0cm}\caption{\label{F1}\small{Contour plots of the joint probability distribution for a two-photon subtracted state  with $\phi=\pi/4$ and a) $r_{A}=r_{B}=0.2$, b) $r_{A}=-r_{B}=0.2$ ($\vert s_{A/B}\vert = 1.74$ dB).}}
\end{figure}

Examples of joint probability distributions $\mathcal{P}(x_{A},x_{B})=\vert \Psi(x_{A},x_{B}) \vert^{2}$ for a two-photon subtracted state with $\phi=\pi/4$ and $r_{A}=r_{B}=0.2$, $r_{A}=-r_{B}=0.2$, are respectively shown in Figure \ref{F1} a) and b).

\section{Ideal detection of non-Gaussian entanglement}\label{sec:IV}

\subsection{Displacement}

A displacement of $\theta$ along the axis $x_{A}=\pm x_{B}$ is produced by the following operator
\begin{equation}\nonumber
\hat{D}(\theta)=e^{-i \theta (\hat{p}_{A} \pm \hat{p}_{B})/2 }.
\end{equation}
The Hamiltonian related to this displacement operator is $H_{\pm}= (\hat{p}_{A} \pm \hat{p}_{B})/2$. The optimal entanglement $E_{Q}$ obtained estimating displacement along the axis $x_{A}=\pm x_{B}$ for a pure two photon-subtracted state with $\phi=\pi/4$ and squeezing parameters $r_A \neq r_B $ is
\begin{equation}\label{Ed}
E_{Q}=\pm e^{r_A + r_B} N(r_A ,r_B) ( \sinh(2 r_A)+ \sinh(2 r_B)+4 \sinh^{2}(r_A)+4 \sinh^{2}(r_B)),
\end{equation}
with 
\begin{equation}\nonumber
N(r_A,r_B)=\frac{32 \sinh(r_A) \sinh(r_B)}{18 + 3 (\cosh(4 r_A)+\cosh(4 r_B))-16(\cosh(2 r_A)+\cosh(2 r_B))+8\cosh(2 r_A)\cosh(2 r_B) + 4 \sinh(2 r_A)\sinh(2 r_B) }.
\end{equation}

\begin{figure}[h]
  \centering
   \subfigure{\includegraphics[width=0.43\textwidth]{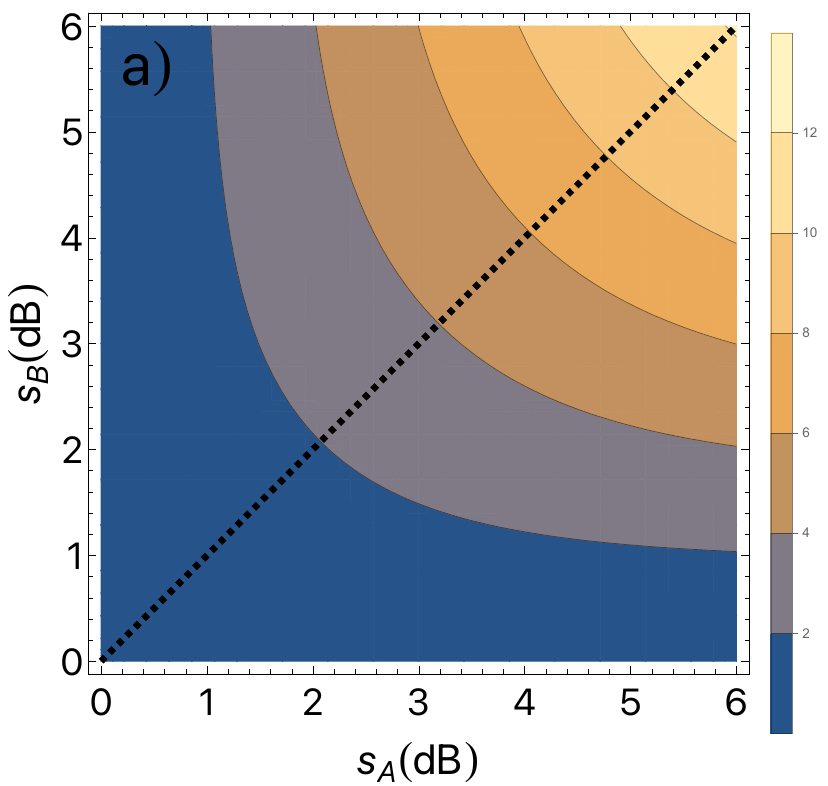}}
    \subfigure{\includegraphics[width=0.45\textwidth]{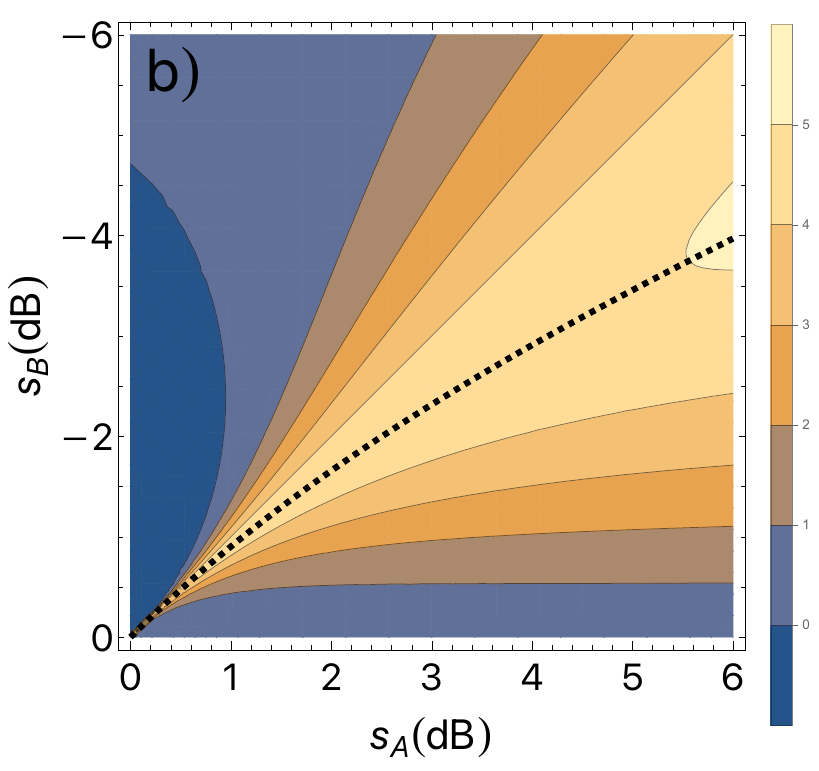}}
\caption{\label{F3}\small{Optimal displacement-estimation entanglement $E_{Q}$ given by Equation (\ref{Ed}) versus squeezing of the input squeezed states $s_A$ and $s_B$. a) Displacement along $x_{A}=x_{B}$ optimizes $E_{Q}$ for states with in-phase input squeezing. b) Displacement along $x_{A}=- x_{B}$ optimizes $E_{Q}$ for states with in-quadrature input squeezing. The black dashed lines in both figures show the optimal choice for $s_A$ and $s_B$ to maximize $E_Q$.}}
\end{figure}

Displacement along either $x_{A}=x_{B}$ or $x_{A}=-x_{B}$ detects entanglement respectively for in-phase squeezing ($r_A, r_B >0$) and in-quadrature --orthogonal-- squeezing ($r_A>0$, $r_B <0$). Figures \ref{F3}a and \ref{F3}b show contour plots of optimal entanglement $E_{Q}$ in the two cases.

\subsection{Phase shift} \label{subsec:ps}
The phase-shift operator 
\begin{equation}\nonumber
\hat{R}(\theta)=e^{-i \theta (\hat{N}_{A} \pm \hat{N}_{B}) } \propto e^{-i \theta (\hat{x}_{A}^2 +\hat{p}_{A}^2 \pm \hat{x}_{B}^2 \pm \hat{p}_{B}^2)/4}
\end{equation}
rotates the state by a phase $\theta$ in local phase subspaces in either clockwise-clockwise ($+$) or clockwise-counterclockwise ($-$) directions. The related Hamiltonian is $H_{\pm}= \hat{N}_{A} \pm \hat{N}_{B}$. The optimal entanglement witness is in this case 
\begin{eqnarray}\label{EdPhase}
E_{Q}=\pm \frac{N(r_A,r_B)^2}{16} (-2+16 \cosh(2 r_A) +
 16 \cosh(2 r_B) - 6\cosh(4 r_A) - 6\cosh(4 r_B) \nonumber \\+ 48 \cosh(4 r_A)\cosh(2 r_B)+48 \cosh(2 r_A) \cosh(4 r_B) -18 \cosh(4 r_A)\cosh(4 r_B)  -84 \cosh(2 r_A) \cosh(2 r_B)  \nonumber \\- 6 \cosh(6 r_A) \cosh(2 r_B) - 6 \cosh(2 r_A) \cosh(6 r_B)+ 34 \sinh(2 r_A) \sinh(2 r_B) \nonumber \\- 3 \sinh(6 r_A) \sinh(2 r_B) - 3 \sinh(2 r_A) \sinh(6 r_B))
\end{eqnarray}

\begin{figure}[h]
  \centering
    \subfigure{\includegraphics[width=0.45\textwidth]{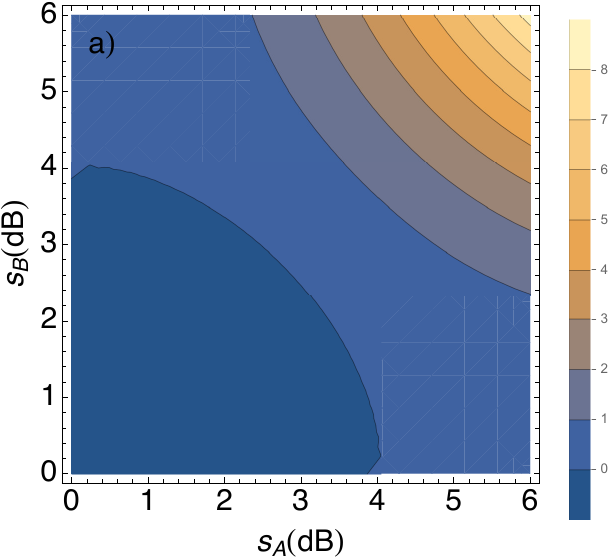}}
    \subfigure{\includegraphics[width=0.47\textwidth]{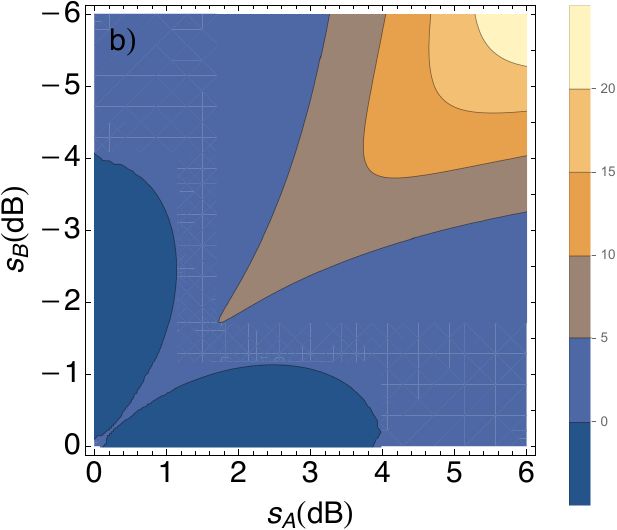}}
\vspace {0cm}\,
\hspace{0cm}\caption{\label{F3a}\small{Optimal phase-estimation entanglement $E_{Q}$ given by Equation \eqref{EdPhase} versus squeezing of the input squeezed states $s_A$ and $s_B$. a) $E_{Q}$ for states with in-phase input squeezing. b) $E_{Q}$ for states with in-quadrature input squeezing.}}
\end{figure}

Entanglement is always detected for clockwise-counterclockwise ($-$) phase shifts, but not for clockwise-clockwise ($+$) as it is just a global phase shift. Figure \ref{F3a} shows contour plots of optimal entanglement $E_{Q}$ for different values of squeezing. Notably, the detection of entanglement does depend on the phase of the input squeezed states unlike the case of one-photon subtraction.

\subsection{Shearing} \label{subsec:sh}
The shearing --also known as phase-gate-- operator 
\begin{equation}\nonumber
\hat{\mathcal{S}}(\theta)=e^{-i \theta (\hat{x}_{A}^2 \pm \hat{x}_{B}^{2})/4 }
\end{equation}
shears the state with respect to the axes $x_{A}$ and $\pm x_{B}$ by a gradient of $\theta$. The related Hamiltonian is $H_{\pm}= (\hat{x}_{A}^2 \pm \hat{x}_{B}^{2})/4$. The optimal entanglement is in this case 
\begin{eqnarray}\label{EdShear}
E_{Q}=\pm \frac{ e^{-2(r_A + r_B)}N(r_A,r_B)^2}{32} (-22 -\cosh(4 r_A) - \cosh(4 r_B) +16 \cosh(2 r_A)  +16 \cosh(2 r_B)   \nonumber \\- 32\sinh(2 r_A) -32\sinh(2 r_B) +16 \sinh(2 r_A) \cosh(2 r_B) +16  \cosh(2 r_A)  \sinh(2 r_B) \nonumber \\+ 4 \sinh(2 r_A) \sinh(2 r_B) - 8 \cosh(2 r_A)  \cosh(2 r_B) + 8\sinh(4 r_A) + 8\sinh(4 r_B)  )
\end{eqnarray}
For two-photon subtraction in the same mode, both shearing with respect to $x_{A}$ and $x_{B}$ and with respect to $x_{A}$ and $-x_{B}$ capture entanglement. Note that in this case the entanglement is maximized for $r_{A/B}<0$, i.e. squeezing along the quadratures $p_{A/B}$, unlike displacement and phase estimation where $E_{Q}$ is maximized for squeezing along $x_{A/B}$.

Figures \ref{F3b}a and \ref{F3b}b show contour plots of optimal entanglement $E_{Q}$ in the cases of input squeezing along the same quadrature (a) or along different quadratures (b).

\begin{figure}[h]
  \centering
    \subfigure{\includegraphics[width=0.45\textwidth]{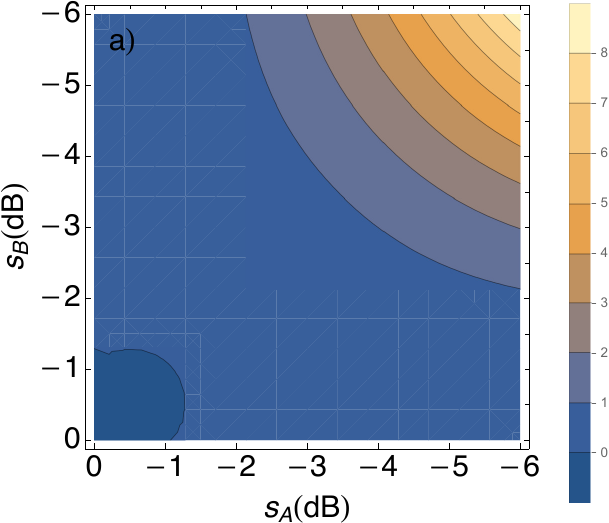}}
    \subfigure{\includegraphics[width=0.45\textwidth]{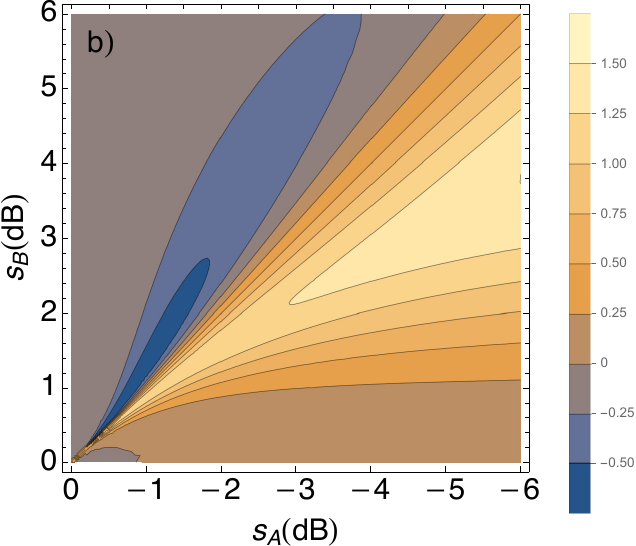}}
\vspace {0cm}\,
\hspace{0cm}\caption{\label{F3b}\small{Optimal shearing-estimation entanglement $E_{Q}$ given by Equation \eqref{EdShear} versus squeezing of the input squeezed states $s_A$ and $s_B$. a) $E_{Q}$ for states with input squeezing along the same quadrature. b) $E_{Q}$ for states with input squeezing along different quadratures.}}
\end{figure}

\begin{figure}[h]
  \centering
    \subfigure{\includegraphics[width=0.45\textwidth]{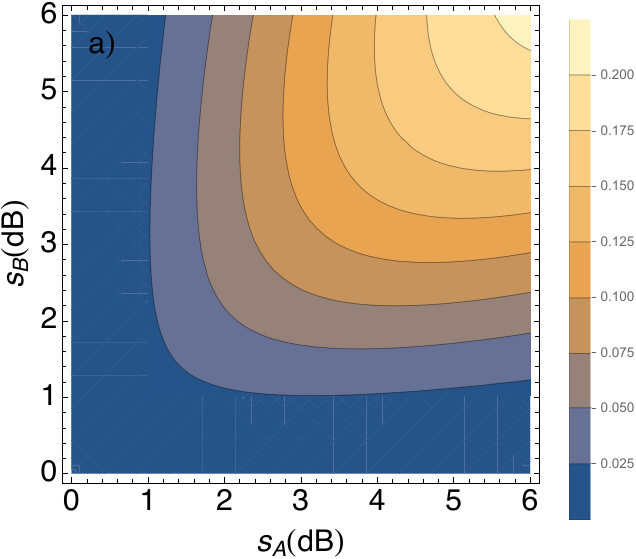}}
    \subfigure{\includegraphics[width=0.45\textwidth]{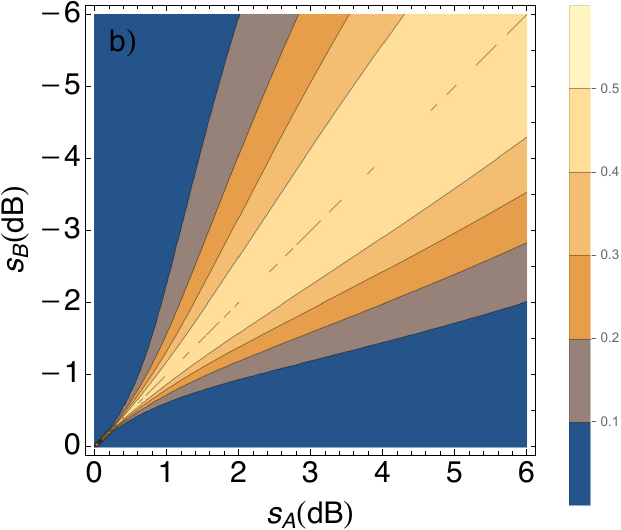}}
\vspace {0cm}\,
\hspace{0cm}\caption{\label{F3b}\small{Optimal squeezing-estimation entanglement $E_{Q}$ given by Equation \eqref{EdSqueezing} versus squeezing of the input squeezed states $s_A$ and $s_B$. a) $E_{Q}$ for states with input squeezing along the same quadrature. b) $E_{Q}$ for states with input squeezing along different quadratures.}}
\end{figure}

\subsection{Squeezing}
The squeezing operator 
\begin{equation}\nonumber
\hat{S}(\theta)=e^{-i \theta (\hat{x}_{A}\hat{p}_{A}+\hat{p}_{A}\hat{x}_{A} \pm \hat{x}_{B}\hat{p}_{B} \pm \hat{p}_{B}\hat{x}_{B})/4 }
\end{equation}
squeezes the position quadratures of modes A and B by a factor of $e^{\theta}$ ($+$) or squeezes the position quadratures of A by $e^{\theta}$ and stretches those of B by $e^{-\theta}$ ($-$). The related Hamiltonian is $H_{\pm}= (\hat{x}_{A}\hat{p}_{A}+\hat{p}_{A}\hat{x}_{A} \pm \hat{x}_{B}\hat{p}_{B} \pm \hat{p}_{B}\hat{x}_{B})/4$. The optimal entanglement is here
\begin{equation}\label{EdSqueezing}
E_{Q}=\pm 4 N(r_A,r_B) \sinh(r_A) \sinh(r_B).
\end{equation}
Interestingly, the joint estimation of the squeezing parameter does detect entanglement unlike for single photon-subtracted states.

\section{Optimal in-quadrature squeezing parameters for displacement-estimation entanglement} \label{sec:V}

The black dashed lines in Figure \ref{F3} a) and b) show the optimal choice of squeezing parameters to maximize the quantity $E_Q$. For in-phase input squeezing, the optimal configuration is simply $s_A = s_B$. On the contrary, and similarly to the one-photon subtracted state case, the maximum of $E$ for in-quadrature input squeezing is not along the diagonal. To estimate the optimal squeezing $s_B<0$ for a given $s_A>0$, we consider the minimal amount of losses $\eta_{\rm min}$ for which the metrological criterion ceases to work for any value of $s_B$ between $-8$ dB and $-s_A$. Going one step backward, i.e., $\eta \lesssim \eta_{\rm min} $, we find the value of $s_B^{\sss \rm opt}$ for which $E$ is maximal (and slightly positive). This value thus corresponds to the state for which the metrological criterion is the most resilient against losses. This is exactly the procedure we use in Figure 9(b) of the main paper. We reproduce below table I of the main paper where optimal $s_B^{\sss \rm opt}$ values can be found for the different $s_A$ squeezing parameters considered in Fig. 9(b) of the paper.
\begin{table}[h!]
    \centering
    \begin{tabular}{|c|c|c|c|c|c|}
     \hline
       $s_A$ (dB)  & 1 & 1.5 & 2 & 2.5 & 3  \\
       \hline
        $s_B^{\sss \rm opt}$ (dB) & -1.4 & -2.6 & -5 & -6.1 & -7.3  \\
         \hline
         $s_B^{\eta = 0}$ (dB) & -0.9 & -1.3 & -1.7 & -2 & -2.3  \\
         \hline
    \end{tabular}
    \caption{Optimal in-quadrature input squeezing parameters $s_B^{\sss \rm opt}$ for two-photon subtracted states. The third row corresponds to the optimal value $s_B^{\eta = 0}$ when $\eta = 0$ (no losses).}
    \label{tab:my_label}
\end{table}

\begin{figure}[h]
  \centering
    \subfigure{\includegraphics[width=0.5\textwidth]{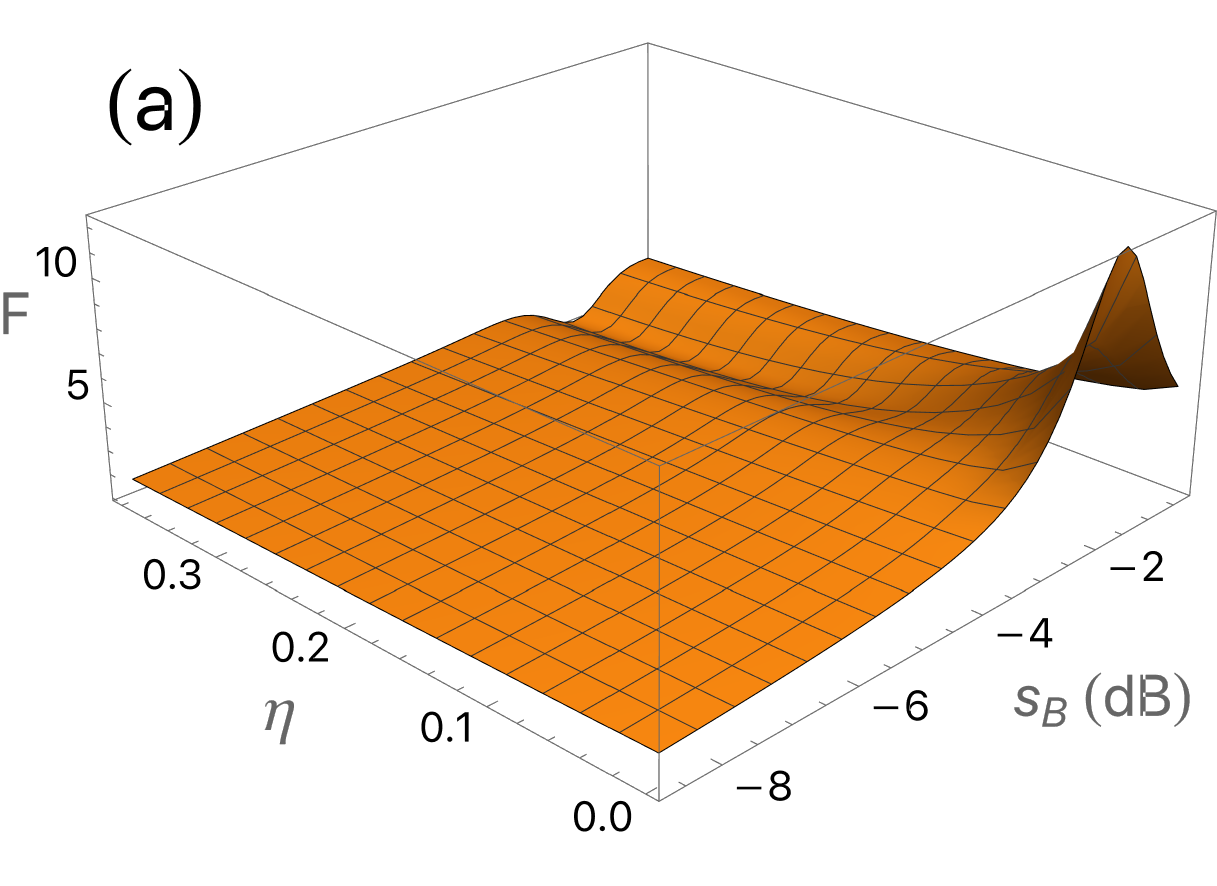}}
    \hspace{1cm}
    \subfigure{\includegraphics[width=0.35\textwidth]{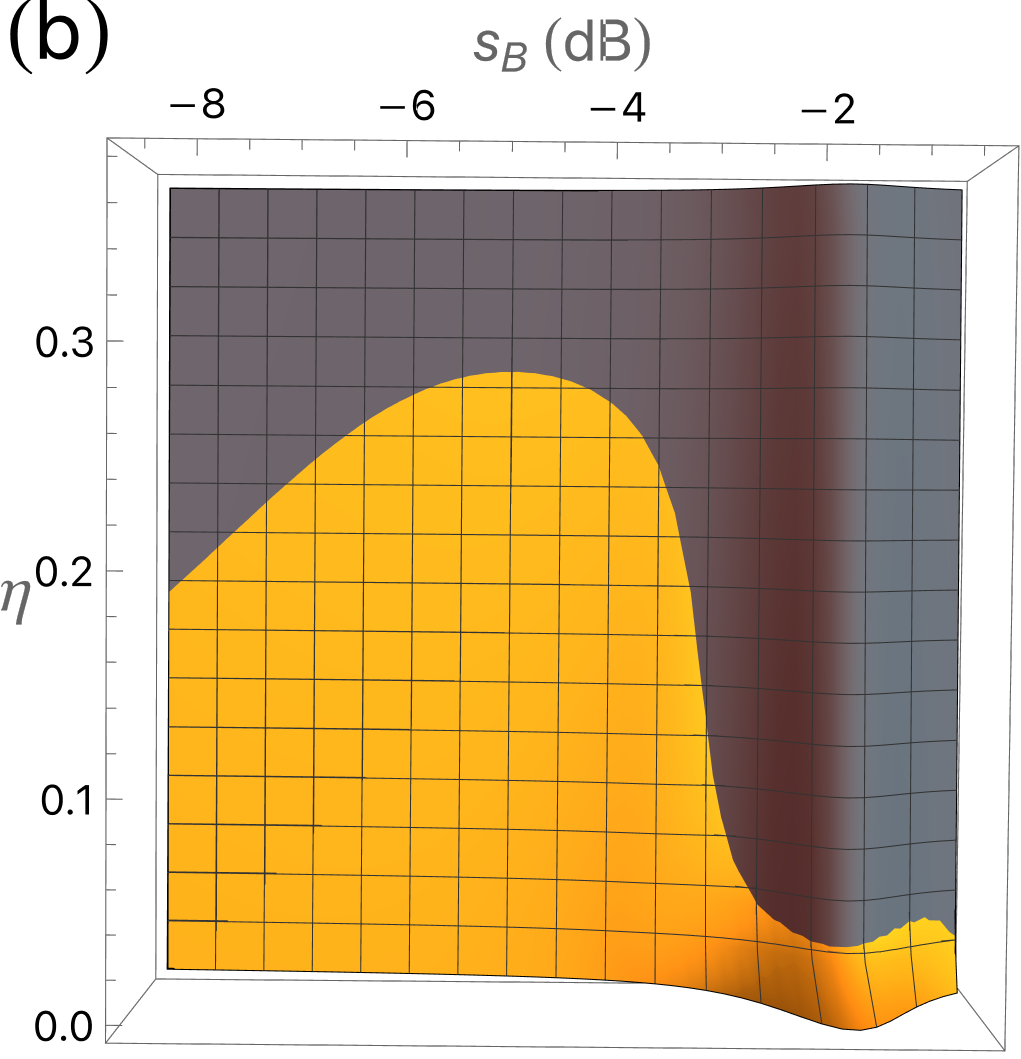}}
\vspace {0cm}\,
\hspace{0cm}\caption{\label{fig:SM6}\small{(a) Fisher information for a two photon subtracted states with $\phi= \pi/4$ and $s_A = 2$ dB as a function of $s_B$ and losses $\eta$. (b) Fisher information (yellow) and local variances (gray) superimposed as a function of $s_B$ and losses $\eta$. The yellow area thus corresponds to $E>0$. }}
\end{figure}

\begin{figure}[h]
  \centering
    \subfigure{\includegraphics[width=0.45\textwidth]{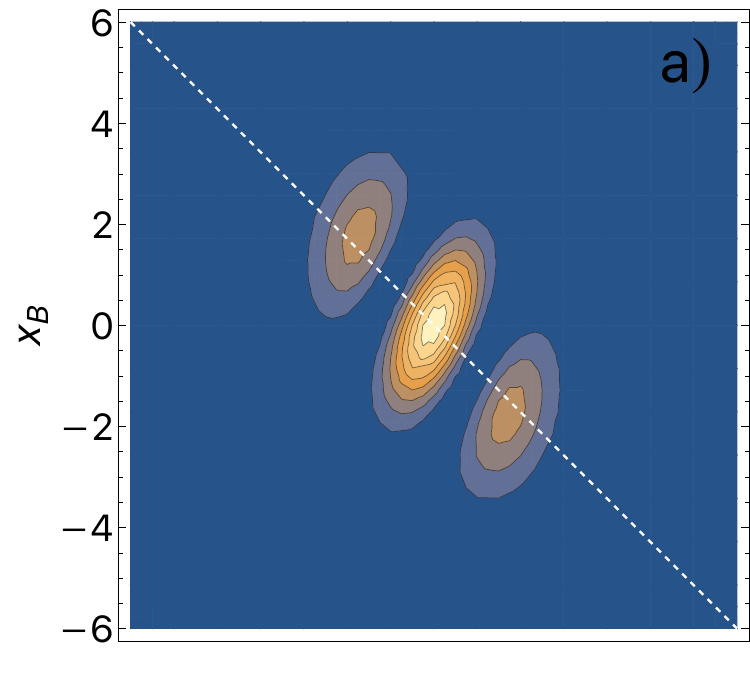}}
    \subfigure{\includegraphics[width=0.413\textwidth]{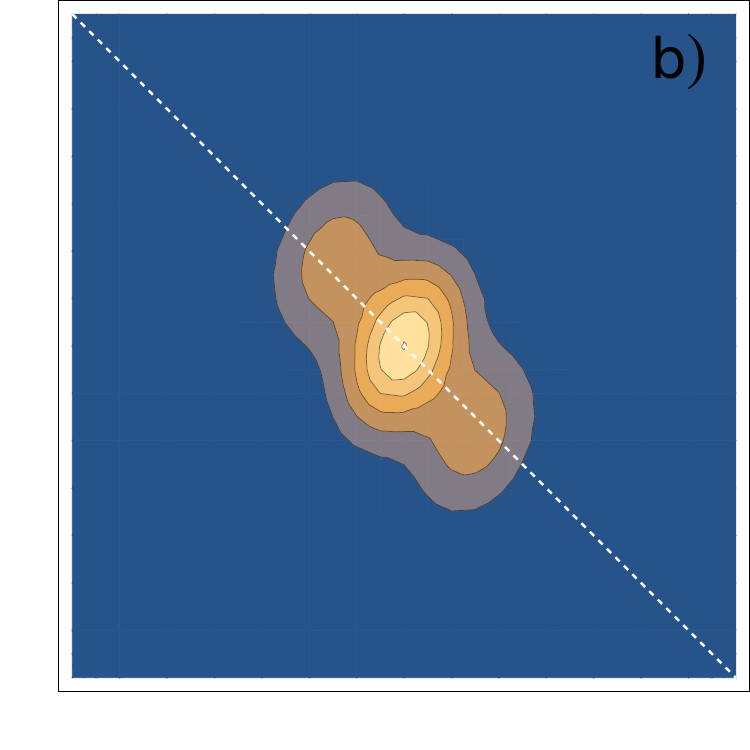}}
       \subfigure{\includegraphics[width=0.45\textwidth]{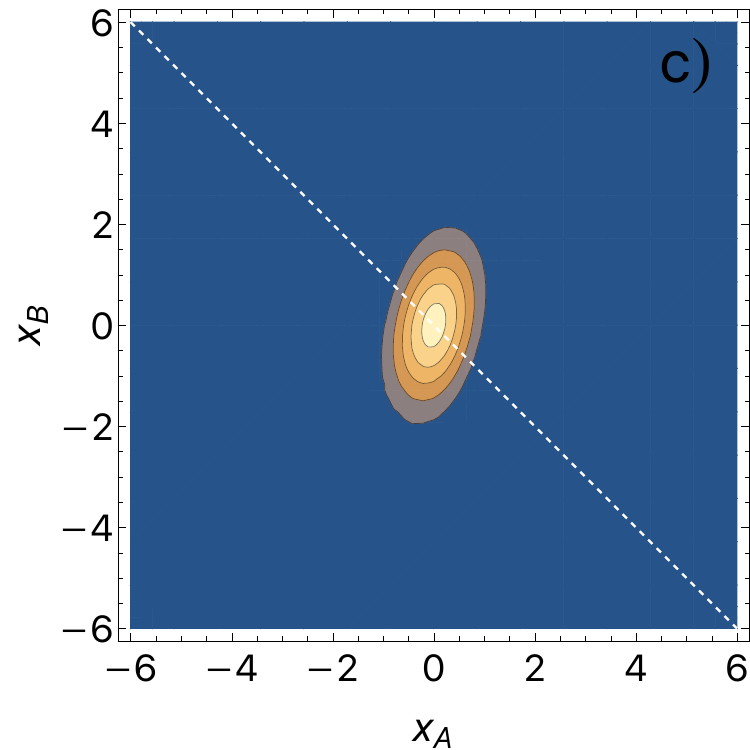}}
    \subfigure{\includegraphics[width=0.415\textwidth]{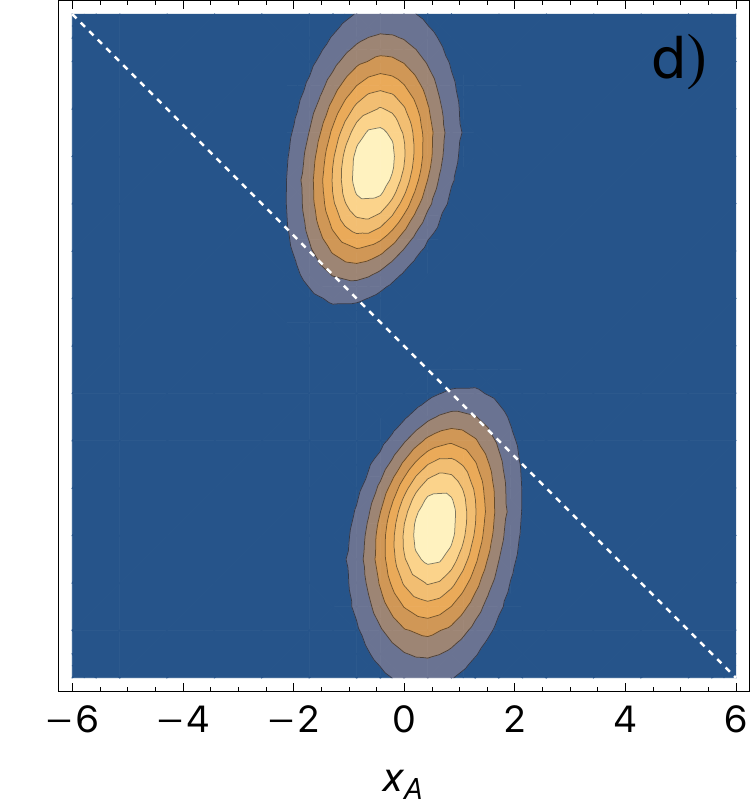}}
\vspace {0cm}\,
\hspace{0cm}\caption{\label{fig:SM7}\small{Joint probability distributions in the $(x_A,x_B)$ plane for a two-photon subtracted state with $\phi =\pi/4$, $s_A = 2$ dB, (a) $s_B = -1.6$ dB, $\eta = 0$ (b) $s_B = -1.6$ dB, $\eta = 30\%$, (c) $s_B = -0.8$ dB, $\eta = 0$, (d) $s_B = -5$ dB, $\eta = 0$.}}
\end{figure}

As shown in Fig. \ref{fig:SM6} (a) for the case $s_A = 2$ dB, the fact that we obtain large values for $|s_B^{\sss \rm opt}|$ results from a better resilience of the Fisher information against losses for larger values of antisqueezing. The peak that we observe in this figure when there is no loss and around $s_B = -1.5$ dB is due to the orientation of the three blobs along the anti-diagonal of the joint probability distribution in the $(x_A,x_B)$ plane [see Fig. \ref{fig:SM7} (a)], because it precisely corresponds to the direction along which we perform our displacement. We also understand easily that due to its richness this quite complicated structure of correlations will be sensitive to any displacement. For squeezing parameters outside this small window the three-blobs-pattern disappears as well as the orientation along the anti-diagonal [see Figs. \ref{fig:SM7} (c) and (d)], resulting in a lower sensitivity to displacement, and thus a lower Fisher information. However, the three-blobs-structure is highly sensitive to losses and quickly fades away when losses increase [see Fig. \ref{fig:SM7} (b)]; this explains in particular the small dip that we observe in Fig. \ref{fig:SM6} for this window of squeezing parameters and for $\eta > 15\%$. On the contrary, the one and two blobs structures of Figs. \ref{fig:SM7} (c) and (d) resist much better to losses. 

When subtracting the variances from the Fisher information to evaluate $E$, we obtain the two-dimensional plot shown in Fig. \ref{fig:SM6}(b) as a function of $s_B$ and $\eta$. The yellow area corresponds to the region where the Fisher information is greater than the sum of local variances, i.e., $E>0$. We clearly see that when $\eta$ increases, larger antisqueezing values resist better to losses.

\section{Variances for the squeezing-based entanglement estimation} \label{sec:VI}

To estimate the metrological entanglement criterion using squeezing operators $\hat H_A = (\hat{x}_A \,\hat{p}_A + \hat{p}_A \,\hat{x}_A)/4$, and $\hat H_B = (\hat{x}_B \,\hat{p}_B + \hat{p}_B \,\hat{x}_B)/4$, one needs to measure the local variances $\text{Var} [\hat H_A]$ and $\text{Var} [\hat H_B]$. This can be done by measuring the kurtosis in four different bases of the quadrature operators: 
\begin{equation}
    \text{Var} \left[\frac{\hat{x}_A \,\hat{p}_A + \hat{p}_A \,\hat{x}_A}{4} \right] = 
    \frac{1}{16} \left[  4 + \frac{4}{3}  \left\langle \left(\frac{\hat{x}_A + \hat{p}_A}{\sqrt2} \right)^4\right\rangle  + \frac{4}{3}  \left\langle \left(\frac{\hat{x}_A - \hat{p}_A}{\sqrt2} \right)^4\right\rangle - \frac{2}{3}  \left\langle \hat{x}_A^4\right\rangle  - \frac{2}{3}  \left\langle \hat{p}_A^4\right\rangle  \right].
\end{equation}